%% file: ttbar_review.tex
\documentclass[epj,nopacs]{svjour}
\pdfoutput=1
\usepackage{color,graphicx,epsfig}
\usepackage{ifpdf}
\usepackage{amsmath}
\usepackage{bm}
\usepackage{color}
\usepackage[english]{babel}
\usepackage{graphicx}%
\usepackage{amsfonts}%
\usepackage{amssymb}
\usepackage{braket}
\usepackage{hyperref}

\definecolor{nicered}{rgb}{0.7,0.1,0.1}
\definecolor{nicegreen}{rgb}{0.1,0.5,0.1}
\hypersetup{colorlinks,citecolor= nicegreen,linkcolor= nicered}

\newcommand{\slashed}{\slash \hspace{-0.19cm}}

\definecolor{Red}{rgb}{1.,0.,0.}

\newcommand{\nn}{\nonumber}
\newcommand{\bea}{\begin{eqnarray}}
\newcommand{\eea}{\end{eqnarray}}
\newcommand{\beq}{\begin{equation}}
\newcommand{\eeq}{\end{equation}}

\newcommand{\be}{\begin{equation}}
\newcommand{\ee}{\end{equation}}

\def\degree{{\ensuremath{^\circ}}}

\newcommand{\AFBtt}{A_{\rm FB}^{t\bar t}}
\newcommand{\AFB}{A_{\rm FB}}
\newcommand{\ASM}{A_{\rm FB}^{\rm SM}}

\newcommand{\Aincl}{A_{\rm FB}^{\rm incl}}
\newcommand{\Alow}{A_{\rm FB}^{\rm low}}
\newcommand{\Ahigh}{A_{\rm FB}^{\rm h}}
\newcommand{\xincl}{\sigma^{\rm incl.}_{t \, \bar{t}}}

\newcommand{\xtt}{\sigma_{t\bar t}}

\begin{document}

\title{Review of new physics effects in $t\bar t$ production}

\author{Jernej F. Kamenik \inst{1}\inst{2} \and Jing Shu\inst{3} \and Jure Zupan\inst{1}\inst{2}\inst{4} 
}                     

\def\LjubljanaFMF{Faculty of Mathematics and Physics, University of Ljubljana,
 Jadranska 19, 1000 Ljubljana, Slovenia }
\def\Cincy{Department of Physics, University of Cincinnati, Cincinnati, Ohio 45221,USA}
\def\LjubljanaIJS{Institut ``Jo\v zef Stefan", Jamova 39, 1000 Ljubljana, Slovenia}
\def\IPMU{Institute for the Physics and Mathematics of the Universe (IPMU), 
the University of Tokyo, Kashiwa, Chiba 277-8568, Japan}

\institute{\LjubljanaIJS  \and 
\LjubljanaFMF \and 
\IPMU \and
\Cincy}

\date{Received: date / Revised version: date}

\abstract{
Both CDF and D\O\ report a forward-backward asymmetry in $t\bar t$ production that is above the standard model prediction. We review new physics models that can give a large forward backward asymmetry in $t\bar t$ production at the Tevatron and the constraints these models face from searches for dijet resonances and contact interactions, from flavor physics and the $t\bar t$ cross section. Expected signals at the LHC are also reviewed.
\PACS{	
      {14.65.Ha}{Top quarks}   \and
      {14.80.-j}{Other particles (including hypothetical)}
     } 
} 

\maketitle
\section{Introduction}
\input{1introduction}

\input{2eft}

\input{3t-channel}

\input{4s-channel}

\input{5lhc}

\input{6conclusions}

\begin{acknowledgement}
This work is supported in part by the European Commission RTN  network, Contract No. MRTN-CT-2006-035482 (FLAVIAnet), by the Slovenian Research Agency, and the World Premier International Research Center Initiative (WPI initiative) MEXT, Japan. J.S. is also supported by the Grant-in-Aid for scientific research (Young Scientists (B) 21740169) from Japan Society for the Promotion of Science (JSPS).
\end{acknowledgement}

 \bibliographystyle{epjc}
 \bibliography{./refs}

\end{document}

%% file: 1introduction.tex
A large forward-backward asymmetry in $t \bar t$ production from $p\bar p$ collisions ($\AFBtt$) has been measured by the CDF and D\O \, collaborations at the Tevatron~\cite{Aaltonen:2011kc,CDF-dilepton,D0EPS}. 
It is significantly larger than the Standard Model (SM) prediction.  We review New Physics (NP) explanations that have been put forward to explain the anomalously large $\AFBtt$. They can be grouped according to whether the main contribution to the NP amplitude is due to $s-$channel or $t-$channel NP particle exchanges. In addition, if the relevant NP degrees of freedom are heavy enough not to be produced in experiments 
they can be integrated out and a general Effective Field Theory (EFT) analysis is possible.

Before we proceed let us first briefly review the experimental evidence for the $t\bar t$ anomaly. CDF reports an inclusive asymmetry $\Aincl = 0.158 \pm0.072 \pm 0.017$
in the $t \, \bar{t}$ rest frame using $5.3 {\rm fb}^{-1}$ of data~\cite{Aaltonen:2011kc} and assuming $m_t = 172.5 \, {\rm GeV}$. Using a channel where both $t$ and $\bar t$ decay semileptonically an even larger asymmetry was found $\Aincl= 0.42 \pm0.15 \pm 0.05$ \cite{CDF-dilepton}. Similarly, a recent D\O\, analysis  using $5.4 {\rm fb}^{-1}$ of data finds $\Aincl = 0.196 \pm0.060^{+0.018}_{-0.026}$~\cite{D0EPS1,D0EPS}.\footnote{Note that D\O \, also reports a leptonic asymmetry $\AFB^l = 0.152\pm0.038^{+0.010}_{-0.013}$ to be compared to MC@NLO prediction of $\AFB^{l,\rm SM} = 0.021\pm0.001$~\cite{D0EPS1,D0EPS}.} We perform a na\"ive weighted average of the three measurements based on independent datasets and combine the statistical and systematic errors in quadrature obtaining $\Aincl = 0.200\pm 0.047$. 
This is to be compared to $\ASM = 0.0724^{+0.0104}_{-0.0067}{}^{+0.0020}_{-0.0027}$ from an approximate NNLO QCD calculation~\cite{Ahrens:2011uf} within SM with $m_t = 173.1 \, {\rm GeV}$ and using MSTW2008 set of PDFs~\cite{Martin:2009iq}.\footnote{In the $p\bar p$ frame, another recent approximate NNLO calculation~\cite{Kidonakis:2011zn} yields $\ASM = 0.052^{+0.000}_{-0.006}$ with $m_t = 173 \, {\rm GeV}$, to be compared with the CDF value of $\Aincl = 0.150 \pm0.058 \pm 0.024$~\cite{Aaltonen:2011kc}. Both SM predictions build upon the recent progress in approximate NNLO calculations~\cite{Moch:2008ai,Czakon:2009zw,Beneke:2009ye,Kidonakis:2008mu,Cacciari:2008zb,Kidonakis:2010dk} and previously known NLO results~\cite{Antunano:2007da,Bowen:2005ap,Kuhn:1998kw}.}  
CDF also reported evidence that the anomalously large asymmetry rises with the invariant mass of the $t \, \bar{t}$ system, with  $\Ahigh\equiv \AFBtt(m_{t\bar t}> 450 {\rm ~ GeV})=0.475\pm0.114$,  while $\Alow\equiv \AFBtt(m_{t\bar t}< 450 {\rm ~ GeV})=-0.116\pm0.153$~\cite{Aaltonen:2011kc}. 
A similar rise of the $\AFBtt$ with the absolute top vs. anti-top rapidity difference $|\Delta y| = |y_t-y_{\bar t}|$ was also reported by CDF with $\AFBtt(| \Delta y | < 1.0) = 0.026 \pm 0.104 \pm 0.056$ and  $\AFBtt(| \Delta y | > 1.0) = 0.611 \pm 0.210 \pm 0.147$~\cite{Aaltonen:2011kc}. The recent D\O\, analysis~\cite{D0EPS}, however, does not observe such rise of the $\AFBtt$ with neither $m_{t\bar t}$ nor $|\Delta y|$. 
For reader's convenience we collect the above results in Table \ref{table:partondata}.

\begin{figure}
\includegraphics[width=0.40\textwidth]{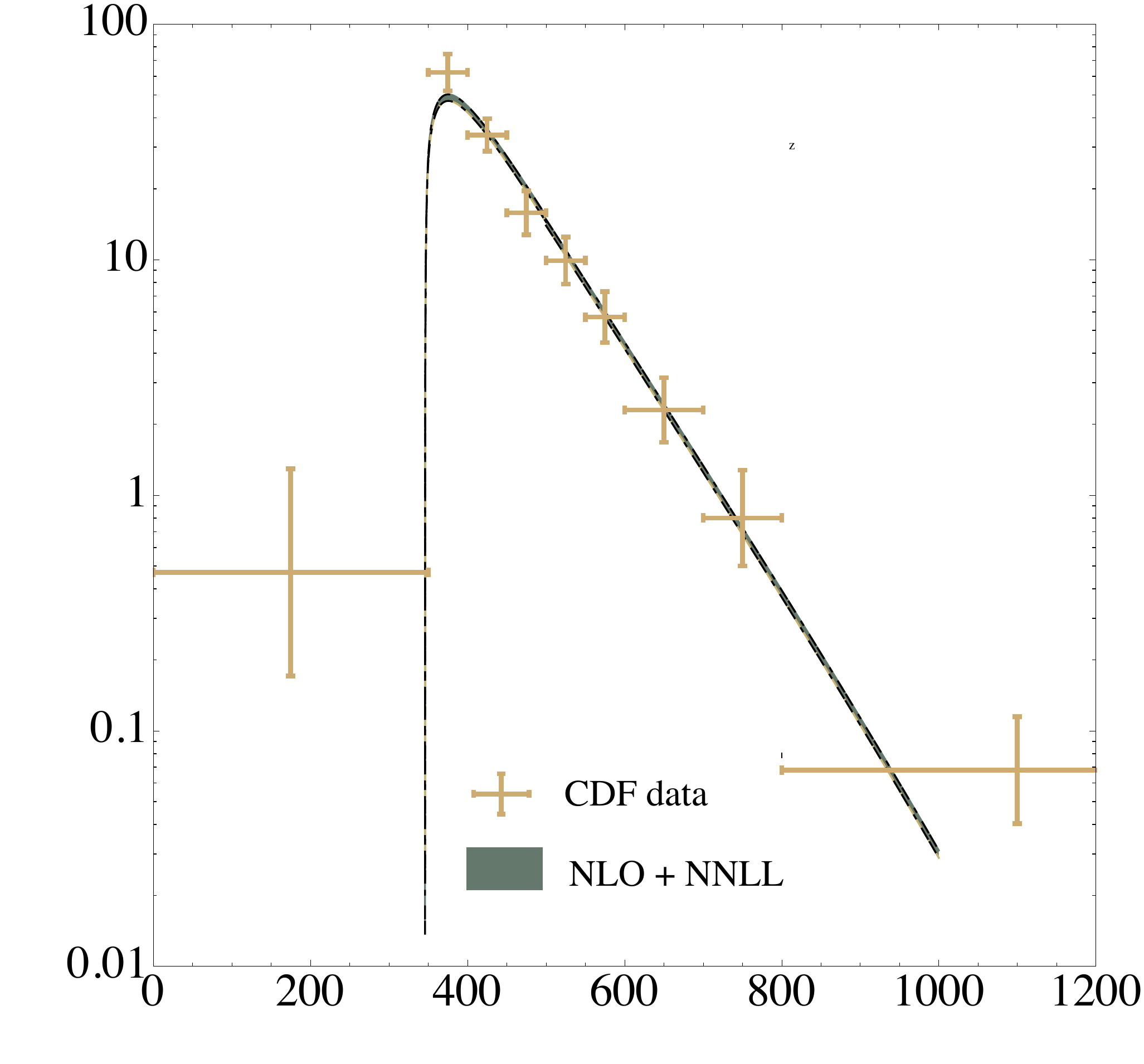}
\caption{\label{fig:CDF-compare} Comparison of the CDF data \cite{Aaltonen:2009iz} for $d\xtt/m_{t\bar t}$ with RG improved QCD prediction at NLO+NNLL order (using the value $m_t=173.1$\,GeV)~\cite{Ahrens:2010zv}. Taken from Ref.~\cite{Ahrens:2010zv}.}
\end{figure}

A very important constraint on NP models that can produce a large $\AFBtt$ is that at the same time they should not significantly affect the $t\bar t$ cross section. Both SM predictions at approximate NNLO  $\sigma^{\rm SM}_{t\bar t}=(6.63^{+0.00}_{-0.27})\,\rm pb$~\cite{Ahrens:2011mw,Ahrens:2011px} at $m_t = 173.1 \, {\rm GeV}$, and $ \sigma_{t \, \bar{t}}^{\rm SM} = 
(7.08^{+0.00}_{-0.24}{}^{+0.36}_{-0.27})\,\rm pb$~\cite{Kidonakis:2011jg} using $m_t = 173 \, {\rm GeV}$ agree well with the measured cross section $\xincl = 6.9 \pm 1.0~{\rm pb}$ from CDF using $4.6 {\rm fb}^{-1}$ of data \cite{Aaltonen:2009iz} and assuming $m_t = 173.1 \, {\rm GeV}$.\footnote{The most recent combination of CDF measurements~\cite{cdfxsec} bears an even smaller error with $\sigma^{\rm incl.}_{t\bar t} = (7.50\pm 0.48)$~pb, but was done assuming $m_t=172.5$~GeV. Using the provided interpolation formulae in~\cite{Ahrens:2011mw} yields $\sigma_{t\bar t}^{\rm SM} = (6.75^{+0.08}_{-0.42})$~pb in the SM at approximate NNLO in QCD.} Good agreement between experiment and SM predictions is
also seen in the differential cross section $d\xtt/d m_{t\bar
  t}$, as shown on Fig. \ref{fig:CDF-compare}.  
  
\begin{figure}
\includegraphics[width=0.40\textwidth]{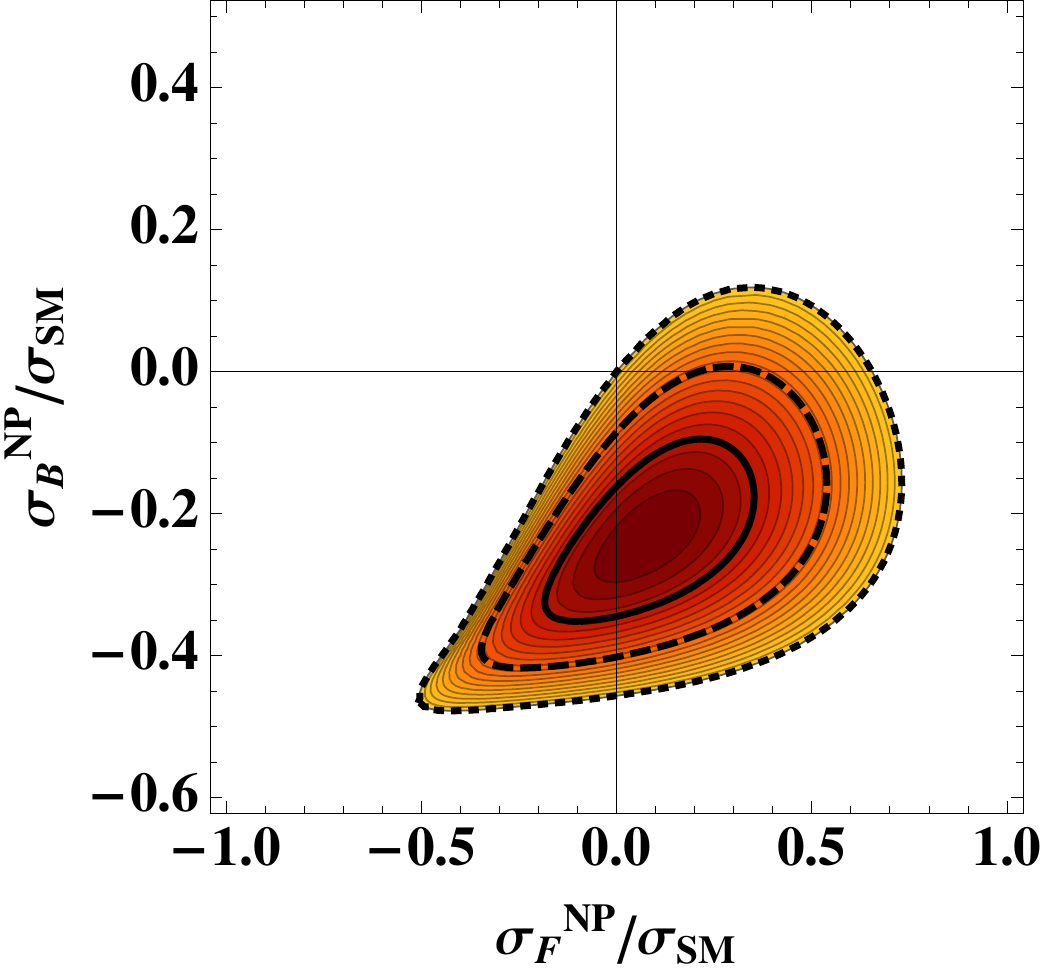}
\caption{$\sigma_F^{\rm NP}$ and $\sigma_B^{\rm NP}$ (normalized to the SM values) needed to explain the measured $\Ahigh$, while at the same time being compatible with $\xtt(m_{t\bar t}>450\,\rm GeV)$. The contours correspond to 1-$\sigma$ (solid), 2-$\sigma$ (dashed) and 3-$\sigma$ (dotted) allowed regions. Taken from Ref.~\cite{Grinstein:2011yv}.}
\label{f.sigmas}
\end{figure}

The constraints these measurements impose on NP models can be expressed in a model independent way \cite{Grinstein:2011yv}.  Let $\sigma_{F,B}^{SM}$ and $\sigma_{F,B}^{NP}$ be the SM and NP forward and backward cross sections, respectively. The latter contain the contributions from NP interfering with the SM and from the NP-matrix elements squared. If interference dominates, $\sigma_{F,B}^{NP}$ can have either sign. If interference is negligible, these terms have to be positive. Using measured and predicted values of  $\AFBtt$ and $\sigma_{t\bar t}$ for $m_{t\bar t}>450$ GeV one obtains the constraints on $\sigma_F^{NP}$, $\sigma_B^{NP}$ shown in Fig. \ref{f.sigmas}.   A preference for $\sigma_B^{NP} <0$ points to an interference effect. If the $s$-channel contribution dominates, this means that the exchanged particle has to be a colour octet vector. The other options are large $t$-channel interference, or a combination of both channels. We shall explore these possibilities in sections \ref{sec:schannel} and \ref{sec:tchannel}, respectively.
 
The paper is organized as follows. In Sec. \ref{EFTsection} we first discuss the model independent implications of the anomalously large $\AFBtt$ measurements as well as the $\xtt$ and $d\xtt/dm_{t\bar t}$ constraints on NP in $t\bar t$ production in the language of EFT. Predominantly $t$-channel models are analyzed in Sec.~\ref{sec:tchannel} while predominantly $s$-channel new physics is discussed in Sec.~\ref{sec:schannel}. In Sec.~\ref{sec:LHC} we discuss the relevant LHC signatures related to NP effects in the $\AFBtt$. We conclude in Sec.~\ref{sec:Conclusions}.

\begin{table*}
\center
\begin{tabular}{ c | c  c } 
\hline \hline 
Observable & Measurement  &  SM predict. \\ 
\hline  
$\Aincl$ & $
  \left.
\begin{matrix}
0.158 \pm 0.072 \pm 0.017\, \text{\cite{Aaltonen:2011kc}}\\
 0.42 \pm0.15 \pm 0.05 \, \text{\cite{CDF-dilepton}}\\
0.196 \pm0.060^{+0.018}_{-0.026} \, \text{\cite{D0EPS}}\\
 \end{matrix}
 \right\} \simeq 0.200\pm0.047
 $ & 
 $(7.24^{+1.04}_{-0.67}{}^{+0.20}_{-0.27})\cdot 10^{-2}  \text{\cite{Ahrens:2011uf}}$\\
$\Ahigh\equiv\AFBtt(m_{t\bar t}>450$GeV)  &  $0.475 \pm 0.101 \pm 0.049$ \cite{Aaltonen:2011kc}  & 
$
(11.1^{+1.7}_{-0.9})\cdot 10^{-2}  \text{\cite{Ahrens:2011uf}}$\\
$\Alow\equiv\AFBtt(m_{t\bar t}<450$GeV) &  $-0.116 \pm 0.146 \pm 0.047$ \cite{Aaltonen:2011kc} & $(5.2^{+0.9}_{-0.6})\cdot 10^{-2} $  \cite{Ahrens:2011uf}\\
$\AFBtt(| \Delta y | < 1.0$) &  $0.026 \pm 0.104 \pm 0.056$  \cite{Aaltonen:2011kc}  & $(4.77^{+0.39}_{-0.35})\cdot 10^{-2} $  \cite{Ahrens:2011uf} \\
$\AFBtt(| \Delta y | > 1.0$) &  $0.611 \pm 0.210 \pm 0.147$  \cite{Aaltonen:2011kc} & $(14.59^{+2.16}_{-1.30})\cdot 10^{-2} $  \cite{Ahrens:2011uf} \\
$\xincl$&  $(6.9 \pm 1.0)$pb \cite{Aaltonen:2009iz}  & 
$\left\{
\begin{matrix}
(6.63^{+0.00}_{-0.27})\text{pb  \cite{Ahrens:2011mw}}  \\ 
(7.08^{+0.00}_{-0.24}{}^{+0.36}_{-0.27})\text{pb  \cite{Kidonakis:2011jg}}
\end{matrix}
\right.
$
\\
\hline\hline
\end{tabular}
\caption{The measurements and predictions of observables in $t\bar t$ production at Tevatron. We quote the approximate NNLO QCD prediction of $\AFB$ from~\cite{Ahrens:2011uf} using MSTW2008 PDFs~\cite{Martin:2009iq}, while the other two choices for PDFs give results in agreement with these~\cite{Ahrens:2011uf}. Among the cross section predictions obtained in~\cite{Ahrens:2011mw} we quote the ${\rm 1PI}_{\rm SCET}$ one, the others being in agreement but with larger quoted errors.}
\label{table:partondata}
\end{table*}

%% file: 2eft.tex
\section{The EFT expansion}
\label{EFTsection}

If the NP degrees of freedom contributing to $t\bar t$ production are heavy enough, they cannot be produced directly in collisions at the Tevatron or the LHC. In term, they can be integrated out and the complete UV theory can be matched onto an Effective Field Theory (EFT) description. All  
the possible NP effects in $t\bar t$ production are then described in terms of effective operators of increasing canonical dimension involving pairs of top and anti-top quarks 
\be
\mathcal L^{t\bar t}_{\rm EFT} =  \sum_{n\geq1}  \mathcal L^{t\bar t}_{n+4}\,. 
\label{eq:LEFT}
\ee
Here the Lagrangian terms $\mathcal L^{t\bar t}_D$ contain contributions from operators of canonical dimension $D$. For simplicity we assume that the operators are composed only of SM degrees of freedom, and in particular contain a $t\bar t$ pair\footnote{For a general discussion on the physics beyond the SM analysis in the EFT approach and classification of all corresponding $D=5,6$ effective operators  c.f.~\cite{Buchmuller:1985jz,Arzt:1994gp,Grzadkowski:2010es}.}. As already stressed, such Wilsonian expansion is applicable to models in which the scale of NP ($\Lambda$) is well above energies probed directly in experiments -- in $t\bar t$ production these can be characterized by $m_{t\bar t}$\,. If the scale $\Lambda$ is too low, in particular if the NP particles can be produced on-shell, the EFT description breaks down. 
Conversely, this also means that as long as the EFT description is valid,  the high-end tail of the $m_{t\bar t}$ spectrum and other observables sensitive to this kinematical region will be affected most by the presence of NP.
This 
can be easily understood by observing that the NP contributions to the $2\to 2$ scattering amplitudes scale as $(E/\Lambda)^{D-4}$, where $E$ is a typical energy scale and $D$ the  dimensionality  of the particular NP operator.
For angularly inclusive observables such as 
$d\xtt/dm_{t\bar t}$,  $\AFBtt$ or charge asymmetries at the LHC there are only two physical energy scales involved, $m_t$ and $m_{t\bar t}$. If therefore $E\sim m_{t\bar t}$, pronounced effects at high $m_{t\bar t}$ are expected.

\begin{figure}
\begin{center}
\vspace{0.5cm}
\includegraphics[width=.45\textwidth]{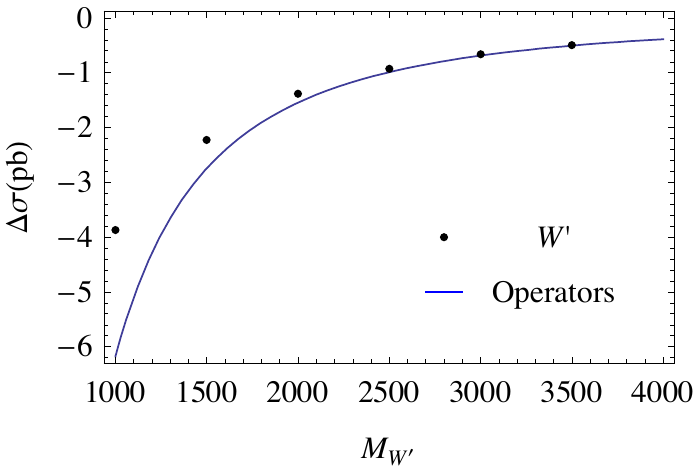}
\vspace{-0.5cm}
\end{center}
\caption{\label{fig:Wp}Correction to the SM cross-section at the LHC due to a $W'$ (whose coupling to $d$ and $t$ quarks is 1) and comparison with the
effective field theory approach. Taken from Ref.~\cite{Degrande:2010kt}.}
\end{figure}

The range of validity of the EFT description at the Tevatron and the LHC has been studied in detail in~\cite{Degrande:2010kt,AguilarSaavedra:2011vw} for a number of models.
In Fig.~\ref{fig:Wp} we show as an example 
a heavy $W'$ contributing in the $t$-channel to the total $t\bar t$ cross-section at the LHC.
The conclusion of  Refs.~\cite{Degrande:2010kt,AguilarSaavedra:2011vw} is that at the Tevatron
the EFT description is an accurate approximation for $t$-channel models provided that the masses ($m_R$) of new exchanged particles satisfy $m_R\gtrsim 1$~TeV. At the LHC, the inclusive cross-section seems to be well approximated provided $m_R\gtrsim 1.5$~TeV. To some extent the same is true even for the $m_{t\bar t}>1$~TeV region, although the effects there tend to be systematically overestimated in the EFT~\cite{AguilarSaavedra:2011vw}. 

The effects of $s$-channel NP are much more dramatic due to the $m_R/\Gamma_R$ resonant enhancement  at on-shell production  (here $\Gamma_R$ is the width of a new resonance and $m_R$ again its mass). Consequently, the Tevatron observables can only be well approximated by EFT if $m_R\gtrsim 1.5$~TeV, while the effects in the $m_{t\bar t}>1$~TeV region at the LHC are grossly underestimated even for significant widths $\Gamma_R\sim 0.1\, m_R$ if $m_R\lesssim 3$~TeV~\cite{AguilarSaavedra:2011vw}\,.

\subsection{Anomalous top-gluon couplings}

Na\"ively, the EFT expansion of (\ref{eq:LEFT}) starts at $D=5$ (or at order $1/\Lambda$) with
\beq
\begin{split}
\mathcal L^{t\bar t}_{5} = -\frac{1}{2}\bar t \big[&(\mu_t + i \gamma_5 d_t )A^{\mu\nu} 
+ (\tilde \mu_t + i \gamma_5 \tilde d_t)g_s T^a G^{a\mu\nu} \\
&+ (\mu'_t+i\gamma_5 d'_t) Z^{\mu\nu}  \big]\sigma_{\mu\nu} t\,,\label{eq:L5}
\end{split}
\eeq
where $\sigma_{\mu\nu} \equiv i [\gamma_\mu,\gamma_\nu]/2$, $T^a$ are the Gell-Mann $SU(3)$ matrices with ${\rm Tr}(T^a T^b) = 2\delta_{ab}$, while $A,Z$ and $G^a$ are the EM, $Z$ and gluon field strength tensors respectively. The coefficients $\mu_t$, $d_t$, $\tilde \mu_t$ and $\tilde d_t$ are the anomalous magnetic, electric, chromomagnetic and the chromoelectric dipole moments of the top quark, respectively. For completeness we have also included the corresponding anomalous $Z$-magnetic and $Z$-electric moments $\mu_t'$ and $d'_t$. All these contributions na\"ively scale as $1/\Lambda$. However, NP with characteristic $\Lambda$ scales much above the EW symmetry breaking scale $v$ ($\Lambda \gg v\sim m_t$) should contribute to the SM action in an EW symmetric way~\cite{Buchmuller:1985jz}. The Lagrangian~(\ref{eq:L5})  then arises from dimension six operators\footnote{A discussion on reducing the overcomplete set of all possible EW symmetric operators can be found in~\cite{Degrande:2010kt,AguilarSaavedra:2008zc}}
\beq
\begin{split}
{\mathcal O}_G &=  \bar Q_3 \phi_u \, \sigma_{\mu\nu} \, g_s T^a G^{a\mu\nu} t_R  \,,\\
 \mathcal O_A &= \bar Q_3 \phi_u \, \sigma_{\mu\nu} A^{\mu\nu} t_R\,,\\
{ \mathcal O}_Z &= \bar Q_3 \phi_u \, \sigma_{\mu\nu} Z^{\mu\nu} t_R\,,
\label{eq:dipoleOperators}
\end{split}
\eeq
where $Q_3=(t_L,d_{3L})^T$ is a doublet of the third-generation quarks  (in the up-quark mass basis), and $\phi_u$ is a scalar field with the quantum numbers of the conjugated SM Higgs doublet, in particular $\braket{\phi_u} = (v,0)^T$\,. 
Denoting
\be
\mathcal L^{t\bar t}_6 = \frac{1}{\Lambda^2}\sum_i c_i \mathcal O_i + \rm h.c.\,,
\label{eq:L6}
\ee
where $c_i$ are dimensionless Wilson coefficients, we can identify~\cite{AguilarSaavedra:2008zc}
\beq
\begin{split}
\{\mu_t,\tilde \mu_t, \mu'_t \}&=- \frac{2v}{\Lambda^2} \{ {\Re}{(c_A)}, {\Re}{(c_G)},{\Re}{(c_Z)}\}\,,\\
 \{d_t , \tilde d_t ,d'_t\} &= -\frac{2v }{\Lambda^2}\{{\Im}{(c_A)}\,,{\Im}{(c_G)}, {\Im}{(c_Z)}\}.
\end{split}
\eeq
In perturbative theories, operators in (\ref{eq:dipoleOperators}) arise only at the one-loop level, leading to a NDA estimate of $c_i/\Lambda^2 \sim g_R^2 / 16 \pi^2 m_R^2$, with  $g_R$ the coupling of the new heavy degrees of freedom to top quarks. Such contributions may nonetheless represent dominant NP effects in models of top compositeness~\cite{Lillie:2007hd}.

\begin{figure}
\includegraphics[width=.47\textwidth]{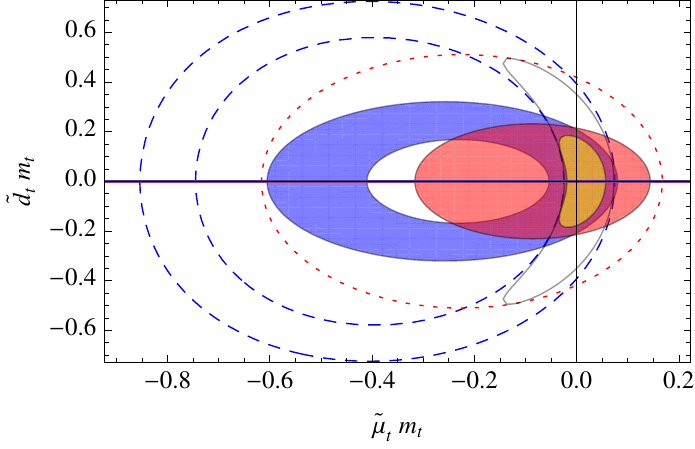}
\caption{\label{fig:cedm} Combined LHC and Tevatron $95\%$ C.L. constraints on the anomalous chromomagnetic ($\tilde \mu_t$) and chromoelectric ($\tilde d_t$) dipole moments (shaded in yellow). Individual constraints come from the total cross-section and $m_{t\bar t}$ spectrum measurements at the Tevatron (dashed blue and doted red), as well as the LHC (shaded blue and red). The combination of only Tevatron constraints is drawn in black. Taken from Ref.~\cite{cedm}.}
\end{figure}

Production of $t\bar t$ pairs at hadron colliders is mostly sensitive to the color octet contributions\footnote{For a general discussion of the CP violating phenomenology associated with operators in (\ref{eq:L5}) c.f.~\cite{Atwood:2000tu}.} associated with $\tilde \mu_t$ and $\tilde d_t$, of which only the CP conserving $\tilde\mu_t$ contribution interferes with the leading SM QCD amplitudes~\cite{Atwood:1994vm,Haberl:1995ek}. 
While none of these contributions can generate a $\AFBtt$ at the Tevatron at leading order in QCD, they would affect both the total cross-section as well as the various kinematical distributions at the Tevatron and the LHC~\cite{Cheung:1995nt,Antipin:2008zx,Gupta:2009wu,Hioki:2009hm,Choudhury:2009wd,Hioki:2010zu,HIOKI:2011xx}. The na\"ive dimension-5 nature of these contributions is reflected in the slower rise of the high $m_{t\bar t}$ spectrum tail, compared to genuine $D=6$ operators~\cite{Blum:2011up}. Nonetheless, a comparison with the Tevatron and LHC data (see Fig.~\ref{fig:cedm}) already constrains the scale $\Lambda$ associated with $\tilde \mu_t$ above $\Lambda>1.1$~TeV (assuming $|\Re(c_G)|=1$)~\cite{cedm}. On the other hand, indirect constraints on the CP violating $\tilde d_t$ contribution are almost two orders of magnitude stronger than the present direct sensitivity of $t\bar t$ production related observables at the Tevatron and the LHC~\cite{cedm}, constraining $\Lambda>5.5$~TeV (for $|\Im(c_G)|=1$).

Additional $D=6$ operators containing top quarks and gluons (i.e. $\bar Q_3 D_\mu \gamma_\nu T^a Q_3 G^{a\mu\nu}$ and $\bar t_R D_\mu \gamma_\nu T^a t_R G^{a\mu\nu}$)  can be matched onto a form factor momentum expansion for the QCD vector and axial quark currents~\cite{Gabrielli:2011jf}
\be
\bar t T^a \gamma_\mu \left[1 + \frac{q^2}{\Lambda^2} (f^t_V + f^t_A \gamma_5)+ \ldots \right]  t G^{a\mu} \,,
\ee
where in $t\bar t$ production $q^2=m_{t\bar t}^2$. The first term in the square brackets is fixed by QCD gauge invariance, while the dots denote higher orders in the $q^2/\Lambda^2$ expansion. Further scalar and pseudoscalar quark density contributions appearing at the same order in $1/\Lambda$ can be reduced to these using equations of motion and gauge invariance. A large $\AFBtt$ at Tevatron can be generated provided an axial current form-factor contribution ($f_A^q$) is present for both the top as well as light valence quarks inside the proton. In particular $f_A^t \simeq f_A^u$ is preferred and with $f_V^q=0$, $f_A^{u,t}=1$ the associated NP scale best accommodating the Tevatron $\AFBtt$ and cross-section measurements is in the range 1~TeV~$< \Lambda < $~1.3~TeV~\cite{Gabrielli:2011jf}\,. Since $f_A^q$ contributions do not interfere with the LO QCD $q\bar q \to t\bar t$ amplitudes, their effects in the cross section measurements at the Tevatron are strictly positive leading to tensions in the high $m_{t\bar t}$ region, especially with the CDF value of $\Ahigh$\, (see also Fig.~\ref{f.sigmas}). Furthermore, since they only contribute at $\mathcal O(1/\Lambda^4)$ these operators in principle compete against $D=8$ NP operators interfering with the SM.

\subsection{Four-quark operators}
\label{sec:4quark}

At $D=6$ we encounter several four-quark operators. 
Among the quark parton luminosities contributing to $t\bar t$ production both at the Tevatron as well as at the LHC, the $\bar u u$ and $\bar d d$ dominate. These are also the only valence contributions that can contribute to $\AFBtt$ at the Tevatron and the charge asymmetries at the LHC. Therefore in the following we focus on flavor conserving operators involving only $u, d$ and $t$ quarks
\beq
\begin{split}
\mathcal O^{q,\hat C}_{\{\rm V,VA\}} &= (\bar q \gamma_\mu \hat C\, q)(\bar t \{\gamma^\mu,\gamma^\mu\gamma_5\} \hat C\, t )\,, \\ 
\mathcal O^{q,\hat C}_{\rm \{AV,A\}} &= (\bar q \gamma_\mu\gamma_5 \hat C\, q)(\bar t\{ \gamma^\mu,\gamma^\mu\gamma_5\} \hat C\, t )\,, \\ 
\mathcal O^{q,\hat C}_{\rm \{S,SP\}} &= (\bar q  \hat C\, q)(\bar t \{1,\gamma_5\} \hat C\, t )\,,\\ 
\label{eq:4quark}
\mathcal O^{q,\hat C}_{\rm\{PS,P\}} &= (\bar q \gamma_5 \hat C\, q)(\bar t \{1,\gamma_5\} \hat C\, t )\,, \\ 
\mathcal O^{q,\hat C}_{T} &= (\bar q  \sigma_{\mu\nu}\hat C\, q)(\bar t \sigma^{\mu\nu} \hat C\, t )\,, 
\end{split}
\eeq
where $q=u,d$, while $\hat C = 1,T^a$ distinguishes between color singlet and octet operators. Possible additional operators can be reduced to these using Fierz identities. Again when matching to particular NP models, an EW symmetric formulation is preferred. However, the number of possible operators is large and the set is highly redundant with respect to $t\bar t$ production phenomenology~\cite{AguilarSaavedra:2010zi}. The transcription between the (axial)vector operators in (\ref{eq:4quark}) and several chiral operator bases in the literature has been provided in~\cite{Degrande:2010kt}.
In perturbative NP realizations such operators can be generated already at the tree level via the exchange of heavy bosonic resonances~\cite{AguilarSaavedra:2011vw} leading to the NDA estimate of $c_i/\Lambda^2 \sim g_R^2/m_R^2$.

Among the operators in (\ref{eq:4quark}) only $\mathcal O^{q,T^a}_{V,A}$ interfere with the SM one gluon exchange amplitudes in $t\bar t$ production\footnote{Here we are neglecting the effects of light quark mass insertions.} and thus contribute to observables already at $\mathcal O(1/\Lambda^2)$. These contributions  have recently been computed to NLO in QCD~\cite{Shao:2011wa}. The remaining operators contribute to $t\bar t$ production at $\mathcal O(1/\Lambda^4)$ and thus in principle compete against $D=8$ operators interfering with the SM (i.e. $\mathcal O^{q,T^a}_{V,A}$ with additional insertions of derivatives). For quantities measured at the Tevatron it has been verified using NDA~\cite{Delaunay:2011gv}, that such higher dimensional contributions are always subleading compared to effects of dimension-six operators. 

The analysis of Tevatron data in presence of $\mathcal O^{q,T^a}_{V,A}$ operator contributions interfering with QCD to $\mathcal O(1/\Lambda^2)$ has been first performed in~\cite{Jung:2009pi,Zhang:2010dr} in terms of chiral operators. The NP effects on $\AFBtt$ and the cross-section to this order are particularly clear in the vector-axial basis (\ref{eq:4quark})~\cite{Degrande:2010kt,Blum:2011up}. Due to parity invariance of QCD it follows that $d\xtt / d {m_{t\bar t}} \propto c^{q,T^a}_{V}$, while $d\AFBtt / d {m_{t\bar t}} \propto c^{q,T^a}_{A}$.
In particular, considering only $\mathcal O_A^{u,T^a}$ contributions to $\mathcal O(1/\Lambda^2)$ the CDF value of $\Ahigh$ requires $c_A^{u,T^a}\simeq (2.4\pm 0.7) (\Lambda/\rm TeV)^2$~\cite{Blum:2011up}, while the bounds on $\mathcal O^{q,T^a}_{V}$ contributions in this approximation from the cross-section and spectrum measurements at the Tevatron can be found in~\cite{Degrande:2010kt}.\footnote{One immediate consequence of these bounds is that the boosted massive jet cross-section as measured by CDF~\cite{CDF10234} cannot be accommodated by EFT contributions to $t\bar t$ final states at $\mathcal O(1/\Lambda^2)$~\cite{Blum:2011up}.}

\begin{figure}
\includegraphics[width=.49\textwidth]{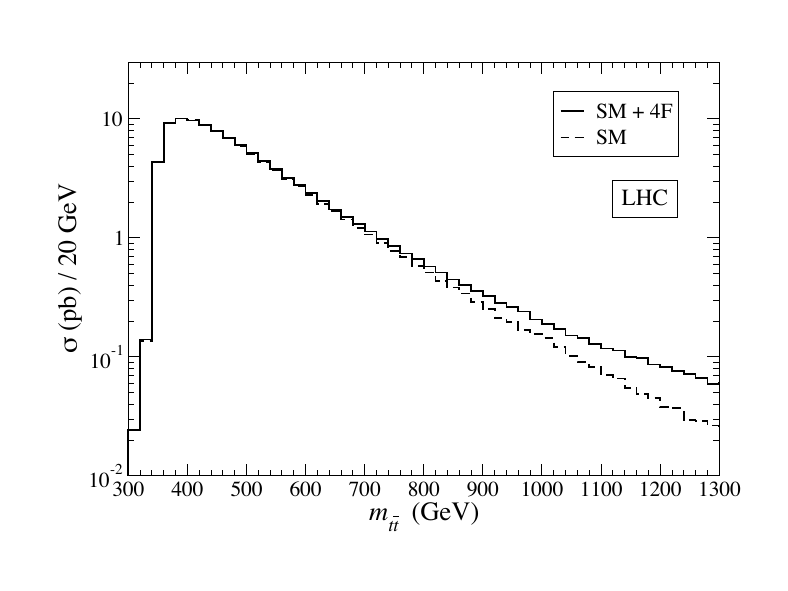}
\vspace{-1.cm}
\caption{\label{mttLHC4f} Invariant mass distribution for $t\bar t$ pairs at LHC, for the SM and with four-fermion contributions accommodating the large $\Ahigh$ as measured by CDF, from Ref.~\cite{AguilarSaavedra:2011vw}.}
\end{figure}

The implication that  purely axial contributions are able to arbitrarily enhance the $\AFBtt$  without introducing associated effects in the cross-section would be misleading. The neglected quadratic NP contributions at $\mathcal O(1/\Lambda^4)$ are necessarily positive definite and (rising as $(m_{t\bar t}/\Lambda)^4$) will tend to enhance the cross-section in the high $m_{t\bar t}$ region~\cite{Blum:2011up}. Such effects have been studied systematically in~\cite{AguilarSaavedra:2011vw} where it was found that including $1/\Lambda^4$ contributions does not spoil the agreement with the Tevatron data, but leads to dramatic effects in the LHC spectrum above $m_{t\bar t}>1$~TeV (see Fig.~\ref{mttLHC4f}).

\begin{figure}
\begin{center}
\vspace{0.5cm}
\includegraphics[width=.4\textwidth]{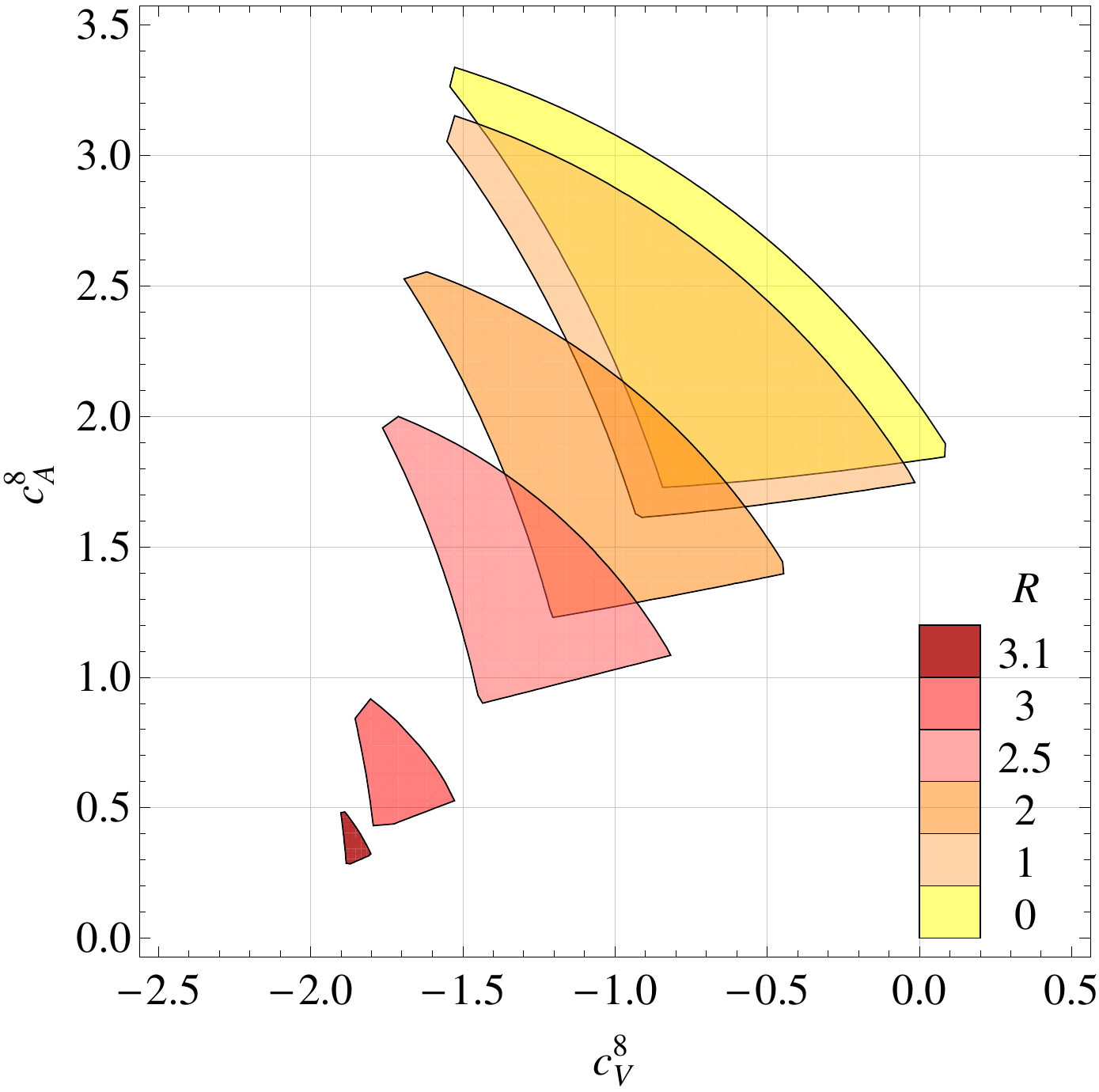}5
\vspace{-0.5cm}
\end{center}
\caption{\label{fig:cVcA}The Tevatron $t\bar t$ observables constraints in the $c^8_V - c^8_A \equiv c^{u T^a}_V - c^{u T^a}_A$ plane: Each region corresponds to the overlap of the $1\sigma$ ranges for $\Ahigh$, $\sigma_{t\bar t}(m_{t\bar t} >450{\rm GeV})$ and $\sigma_{t\bar t}(700{\rm GeV} < m_{t\bar t} < 800{\rm GeV})$ for different
values of $R$ in eq.~(\ref{eq:R}). Taken from Ref.~\cite{Delaunay:2011gv}.}
\end{figure}

The consistency of the Tevatron measurements at high $m_{t\bar t}$ in presence of all the operators in (\ref{eq:4quark}) and their contributions to $\mathcal O(1/\Lambda^4)$ has been analyzed in~\cite{Delaunay:2011gv}. For this purpose it is useful to parametrize the contributions of the operators not interfering with the SM as
\be
R^2 \equiv w_+^2 + w_-^2 + r_{ST}^2 + r_P^2, 
\label{eq:R}
\ee
where $w_{\pm}$, $r_{ST}$ and $r_P$ are quadratic functions of $\{c^{q,\hat C}_{AV},$ $c^{q,\hat C}_{VA}, c^{q,1}_{V}, c^{q,1}_A\}$, $\{c^{q,\hat C}_S,c^{q,\hat C}_T\}$ and $\{ c^{q,\hat C}_P, c^{q,\hat C}_{SP}, c^{q,\hat C}_{PS} \}$ respectively, and have been defined in~\cite{Delaunay:2011gv}.  Then, depending on the value of $R$, the Tevatron measurements of $\Ahigh$, $\sigma_{t\bar t}(m_{t\bar t} >450{\rm GeV})$ and $\sigma_{t\bar t}(700{\rm GeV} < m_{t\bar t} < 800{\rm GeV})$ can all be accommodated within one standard deviation in certain regions of the $(c^{u T^a}_V,c^{u T^a}_A)\equiv(c_V^8,c_A^8)$ plane at $\Lambda=1$~TeV as shown in Fig.~\ref{fig:cVcA}\,. 

A number of conclusions can be reached: i) the size of non-interfering contributions is bounded from above by $R_{\rm max}\simeq 3.1$; ii) individual purely chiral contributions (i.e. $c^{u T^a}_V = \pm c^{u T^a}_A$) cannot accommodate all the measurements -- in particular the large measured $\Ahigh$ values cannot be reproduced; iii) a minimal axial-vector contribution of $c^{u T^a}_A\gtrsim 0.3$ is required; iv) a pure axial-vector contribution $c^{u T^a}_A\simeq 1.8$ at $R\simeq0$ is consistent with the Tevatron data; (5) finally, the $1/\Lambda^4$ contributions are able to reproduce the CDF boosted top cross-section result~\cite{CDF10234}, while remaining in agreement with the inclusive cross-section and $m_{t\bar t}$ spectrum measurements.  However, also in this most general scenario accommodating the central value of $\Ahigh$ would imply a significant enhancement in the LHC $t\bar t$ cross-section at high $m_{t\bar t}$, something that the early ATLAS results on the ${t\bar t}$ cross-section above $m_{t\bar t}>1$~TeV~\cite{cedm,ATLAS-CONF-2011-087} do not indicate.

%% file: 3t-channel.tex
\section{Predominantly $t$-channel models}
\label{sec:tchannel}
In this section we review models that can produce a large $\AFBtt$ with NP states that are light, ${\mathcal O}(200-600)$ GeV, and 
contribute predominantly in the $t-$channel. The models face a number of tight constraints, which make their structure highly nontrivial. While giving a large $\AFBtt$ the models at the same time should not significant affect the $d \xtt/m_{t\bar t}$, at least below $m_{t\bar t}\lesssim 1$ TeV. They should obey constraints from dijets, not lead to excessive production of same sign tops, should not modify the single top production cross section too much and satisfy stringent flavour constraints. 

The ``$t-$channel" models that have been invented to explain a large  $\AFBtt$ can be grouped into three categories: i) models with large flavour violation, ii) flavour conserving models, and iii) models that do not lead to $t\bar t$ final state, but to related final states $t\bar t+X$ (this also means there is no interference with the SM $t\bar t$ production).

\begin{figure}
\includegraphics[width=0.45\textwidth]{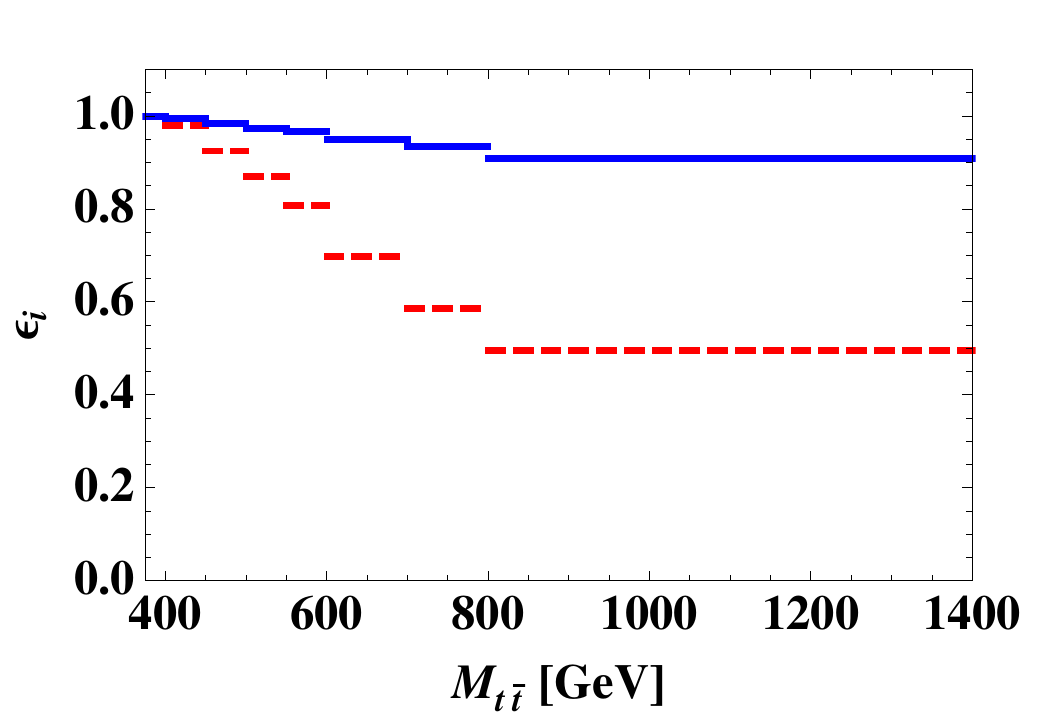}%
\caption{Acceptance correction factors $\epsilon_i$ in (\ref{eq:epsi}) for a model with a vector in an octet representation of $SU(3)_U$ flavour group. Examples for 300 GeV (dashed red) and 1200 GeV (solid blue) are shown. Taken from Ref.~\cite{Grinstein:2011to-appear}.}\label{Fig-eff}
\end{figure}

Before we review the models, let us make a minor detour to explain a technical detail, which however, can be important in judging the viability of models. In their analyses, CDF also quotes $\AFBtt$ and $d\xtt/dm_{t\bar t}$ ``deconvoluted" to ``partonic" or ``truth" level~\cite{Aaltonen:2011kc,Aaltonen:2009iz}. These quantities may be the easiest for theorists to compare with what a particular NP model predicts, avoiding the need to perform detector simulations. However, there is a caveat in that the deconvolution was done assuming SM $t\bar t$ production. In the limit of infinitely small bins in $m_{t\bar t}$, $\Delta y$, and for $4\pi$ detector coverage this would have no effect, but with finite bins there is an error associated with the deconvolution. Especially for very forward $t\bar t$ production this may be a problem as CDF's acceptance for semileptonic tops drops quickly in the rapidity region $|y|\gtrsim 1$. This can have a significant effect on the measured $d\xtt/dm_{t\bar t}$ in the high $m_{t\bar t}$ bins as pointed out in Refs.~\cite{Gresham:2011pa,Jung:2011zv}. A prescription of how to quickly estimate the effects of deconvolutions when scanning over large parameter spaces can be found in~\cite{Grinstein:2011to-appear}. The Òcorrection factorsÓ to be used when comparing with CDF $d\xtt/dm_{t\bar t}$ measurement are shown in Fig.~\ref{Fig-eff} for an example vector model, where 
\beq
\left(\frac{d\sigma^{\rm NP}}{dm_{t\bar t}}\right)^{CDF}_i=\epsilon_i \times \left(\frac{d\sigma^{\rm NP}}{dm_{t\bar t}}\right)_i\,.
\label{eq:epsi}
\eeq
Finally we not note that these effects are much less important at the LHC, due to the higher pseudo-rapidity coverage of the ATLAS and CMS detectors~\cite{Gresham-t}.

\subsection{Large Flavor Violation}
A very common ingredient in models that give a large $\AFBtt$ is the presence of large non-universal flavour violation in the sense that new states couple strongly to $u-t$ or $d-t$ flavour changing quark currents but only weakly or not at all to $u-c$ \, or \, $c-t$~\cite{Jung:2011zv,Jung:2009jz,Cheung:2009ch,Frampton:2009rk,Shu:2009xf,Arhrib:2009hu,Dorsner:2009mq,Cao:2009uz,Barger:2010mw,Cao:2010zb,Xiao:2010hm,Cheung:2011qa,Cao:2011ew,Shelton:2011hq,Berger:2011ua,Barger:2011ih,Bhattacherjee:2011nr,Patel:2011eh,Barreto:2011au,Craig:2011an,Buckley:2011vc,Shu:2011au,Jung:2011ua,Fox:2011qd,Fox:2011qd,Cui:2011xy,Duraisamy:2011pt,AguilarSaavedra:2011ug,Blum:2011fa}\,. While it is possible to arrange couplings in this way in concrete models, this may be hard to achieve  without tunings~\cite{Chivukula:2010fk,Bai:2011ed}, or conversely without tunings the asymmetry would be SM-like as in Randall-Sundrum (RS) models of flavour~\cite{Bauer:2010iq}. As an example let us consider the non-Abelian model of Ref.~\cite{Jung:2011zv}. Here $t_R$ ad $u_R$ constitute a doublet of a non-Abelian horizontal gauge group $SU(2)_X$, while all the remaining quark fields including $c_R$ are not charged under this group. The gauge bosons of the $SU(2)_X$ are dubbed $W'^{\pm}$ and $Z'$ and are EM neutral (also the $W'^{\pm}$). The non-Abelian nature of horizontal interactions helps to avoid the same sign top pair production constraints, that exclude the original Abelian model~\cite{Jung:2009jz}.

If $SU(2)_X$ is broken by a scalar doublet, then one has a $SU(2)_X$ custodial symmetry and $m_{W'}=m_Z'$. For a viable phenomenology this custodial symmetry needs to be broken. The non-custodial Lagrangian is~\cite{Jung:2009jz}
\beq
\begin{split}
{\cal L_{\rm \slashed{cust.}}} &= \frac{g_X}{\sqrt{2}} W'^{-}_\mu \Big\{ \bar{t}_R \gamma^\mu t_R (-cs) \,+\, \bar{u}_R \gamma^\mu u_R (cs) \\
&\,+\, \bar{t}_R \gamma^\mu u_R (c^2) \,+\, \bar{u}_R \gamma^\mu t_R (-s^2) \Big \} \\
& +\, \frac{g_X}{\sqrt{2}} W'^{+}_\mu \Big\{ \bar{t}_R \gamma^\mu t_R (-cs) \,+\, \bar{u}_R \gamma^\mu u_R (cs) \\
&\,+\, \bar{t}_R \gamma^\mu u_R (-s^2) \,+\, \bar{u}_R \gamma^\mu t_R (c^2) \Big \} \\
& +\, \frac{g_X}{2} Z^\prime_\mu \Big\{ \bar{t}_R \gamma^\mu t_R (c^2-s^2) \,+\, \bar{u}_R \gamma^\mu u_R (s^2-c^2)\\
& \,+\, \bar{t}_R \gamma^\mu u_R (2cs) \,+\, \bar{u}_R \gamma^\mu t_R (2cs) \Big\}.
\label{interaction}
\end{split}
\eeq
where $c=\cos \theta$, $s=\sin \theta$ and $\theta$ parametrizes the mismatch between the $(u_R,t_R)$ gauge and mass eigenstates. For $\theta\ne 0$ the $t_R$ and $u_R$ flavour numbers are broken. On the one hand $\cos \theta$ needs to be close to one in order to get a large $\AFBtt$ from the $\bar u_R \slashed W'{}^+ t_R$ coupling. At the same time, a large $\cos\theta$ (in particular $\cos\theta>0.92$)  is required by di-jet constraints so that $\bar u_R\slashed W'{}^\pm u_R$ couplings are sufficiently reduced. On the other hand, suppressing di-jet production through the $\bar u_R \slashed Z' u_R$ coupling requires $\cos \theta<1$.
The prefered choice of Ref. \cite{Jung:2011zv} is a parameter choice "A" with $m_{W'}=200$ GeV, $m_{Z'}=280$ GeV, $\alpha_X=0.060$ and $\cos\theta=0.95$, which gives  $\Ahigh=0.22$ (0.30 if acceptance corrections factors would not have been taken into account). There are also extra states ensuring that $SU(2)_X$ is not anomalous, but they are not needed in the low energy $t\bar t$ phenomenology. 

Let us next discuss the flavour structure of the theory in more detail (another example of such large violation in the right-handed sector is given in Ref.~\cite{Shelton:2011hq}). Generating the required flavour patterns seems to be challenging within a concrete model of flavour. For instance, if there is an $SU(2)_X$ doublet that obtains a vacuum expectation value, this can generate the top mass from a dimension 5 operator
\beq
 {\cal L} \ni  \frac{(\lambda_u^\prime)_i}{M} \,( \bar{Q}_i \cdot h_{SM} )\, (\phi_D \cdot q)
\eeq
with $q=(t_R,u_R)$. The scale $M$ cannot be much above the electroweak scale $v\simeq m_t$ in order to obtain the observed large top mass. How one obtains the $c$- and $u$-quark masses is not specified in~\cite{Shelton:2011hq}, but one option is that the charm quark mass is generated from a dimension 4 operator (the SM Yukawa term), while the up quark mass comes again from higher order operators. There is an immediate vacuum alignment problem.  The charm quark direction needs to be aligned finely so that no $\bar u_r \slashed W' c_R$ and $\bar u_R\slashed Z' c_R$ couplings are present. Furthermore, the directions of the scalar vacuum expectation values giving masses to $W'$, $Z'$ need to be aligned with the top quark mass direction at the level of $\sim 5\%$. Note that these are necessarily different scalars since the $SU(2)_X$ custodial symmetry needs to be broken. 

The above alignment or tuning of interactions in flavour space is common to models with large flavour violation. As another example let us mention a non-supersymmetric SU(5) GUT model~\cite{Dorsner:2009mq,Dorsner:2010cu}. The part relevant for the $t\bar t$ phenomenology is the 45-dimensional Higgs representation that is split. There are two light scalars with TeV scale masses with the following $SU(3)_c\times SU(2)_L\times U(1)_Y$ quantum numbers, $\Delta_6\sim (3,1,-4/3)$ and $\Delta_1\sim (8,2,1/2)$. These scalars do not mediate proton decay at the tree level, while the remaining part of the multiplet is heavy. To have gauge coupling unification there is a similar split in a 24-dimensional fermionic multiplet that also gives neutrino masses through a combination of type I and III see-saws. To have a positive $\AFBtt$ one needs $m(\Delta_6)\sim 300$ GeV. At the same time $\Delta_1$ needs to be heavier, $m(\Delta_1)\sim 1$ TeV, as its contribution to $\AFBtt$ is negative. 

Note that as is typical for models where large $\AFBtt$ is linked to large flavour violation, also in the GUT model of~\cite{Dorsner:2009mq,Dorsner:2010cu} the couplings of $\Delta_{1,6}$ to fermions must have a very constrained structure. We restrict our discussion to $\Delta_6 - u_i - u_j$ interactions
\beq
\mathcal L_{\Delta_6} \ni \frac{g^{ij}_6}{2} \epsilon_{abc} \bar u_{R i}^a {u_{R j}^{C\,b}} \Delta_6^c + \rm h.c.\,,
\eeq
where $a,b,c$ denote colour and $i,j$ flavour indices, while $\epsilon_{abc}$ is a completely antisymmetric tensor with $\epsilon_{123}=1$. In order to explain the measured $\AFBtt$ the $u-t-\Delta_6$ coupling needs to be large
\beq
|g_6^{13}|=0.9(2)+2.5(4)\frac{m_{\Delta_6}}{1~{\rm TeV}}.
\eeq
On the other hand, although the model avoids Flavour Changing Neutral Current (FCNC) constraints at the tree-level by the virtue of the antisymmetric nature of $g_6^{ij}=-g_6^{ji}$, the one loop contributions to $D-\bar D$ mixing require  the $c-t-\Delta_{6}$ coupling to be much smaller,  $|g_6^{23}|<0.0038$. The di-jet and single top production at the Tevatron furthermore put bounds on the $u-c-\Delta_{6}$ coupling, with $|g_6^{12}|\lesssim 0.03$ for $m_{\Delta_6}=300$ GeV. Such a hierarchy of couplings to di-quarks, where they couple most strongly to the 1st and 3rd generation was dubbed "perverted" in \cite{Giudice:2011ak}.

Apart from the required non-trivial flavor structures, it has been recently pointed out~\cite{Gresham:2011pa,Blum:2011fa} that despite the acceptance effects in~\eqref{eq:epsi} most of the $t$-channel models with new flavor violating scalars cannot reproduce the CDF measurement of $\Ahigh$ without being in conflict with the $d\xtt/d m_{t\bar t}$ measurements at high $m_{t\bar t}$. In particular, a colour singlet $SU(2)_L$ doublet with hypercharge $Y=-1/2$ was singled out in~\cite{Blum:2011fa} as the only representation able to accommodate all CDF measurements within $1~\sigma$ provided its mass lies below $m_R\lesssim 250$~GeV.

\subsection{Flavor conserving models}
The $t\bar t$ production is not flavour violating. In the SM for instance it proceeds predominantly through a single gluon exchange. The $p\bar p$ initial state does not carry a nonzero net flavour number, and nor does the $t\bar t$ final state. So why even consider NP models with large flavour violation in order to explain the anomalous $\AFBtt$? Let us briefly consider $s$-channel dominated NP models first. In order to have $\AFBtt >0$ such NP needs to couple to $q\bar q$ and $t\bar t$ with opposite signs~\cite{Cao:2010zb,Bai:2011ed}. The required couplings are thus flavour diagonal, but they are not flavour universal! This means that there is an inherent flavour violation in the $s$-channel models and FCNCs are likely to be generated, unless couplings are tuned to be exactly diagonal.

In $t$-channel models one needs large $u-t$ or $d-t$ couplings in order to have sizable contributions to $t\bar t$ production starting from $p\bar p$ initial state. As already discussed,  in concrete models one then has to  worry about FCNCs~\cite{Dorsner:2010cu,Shelton:2011hq}. There are two options to deal with these. The first is to make the couplings of new states to $c-t$ and $u-c$ small \cite{Jung:2011zv,Jung:2009jz,Gresham:2011dg}. The second possibility is that the FCNCs are small because the models are protected by flavour symmetries~\cite{Grinstein:2011yv,Delaunay:2011vv,Babu:2011yw,Ligeti:2011vt,Tavares:2011zg}.

We now discuss the second possibility. To start with, we perform a quick general counting of possible flavour models. For simplicity, let us assume that in the leading approximation NP respects the SM flavour symmetry group $G_F=SU(3)_U\times SU(3)_D\times SU(3)_Q$, which is a global symmetry of the SM quark sector if the Yukawas are set to zero. Listing all possible scalar and vector fields that can couple to quarks through renormalizable interactions respecting $G_F$ gives 20 possible charge assignments for vectors and 16 for scalars. Most of these could contribute to the $\AFBtt$. A complete analysis can be found in \cite{Grinstein:2011to-appear}, while here we just quote results for a vector that is both an octet of colour and and an octet of $SU(3)_U$, as it was also presented in~\cite{Grinstein:2011yv}. To compute the size of FCNCs one needs to also specify the size of flavour breaking. For concreteness one can assume Minimal Flavour Violation (MFV) (c.f.~\cite{D'Ambrosio:2002ex,Kagan:2009bn}), but this is not really essential, as long as the breaking is small for the first two generations. 

\begin{figure}[t]
\includegraphics[width=0.475\textwidth]{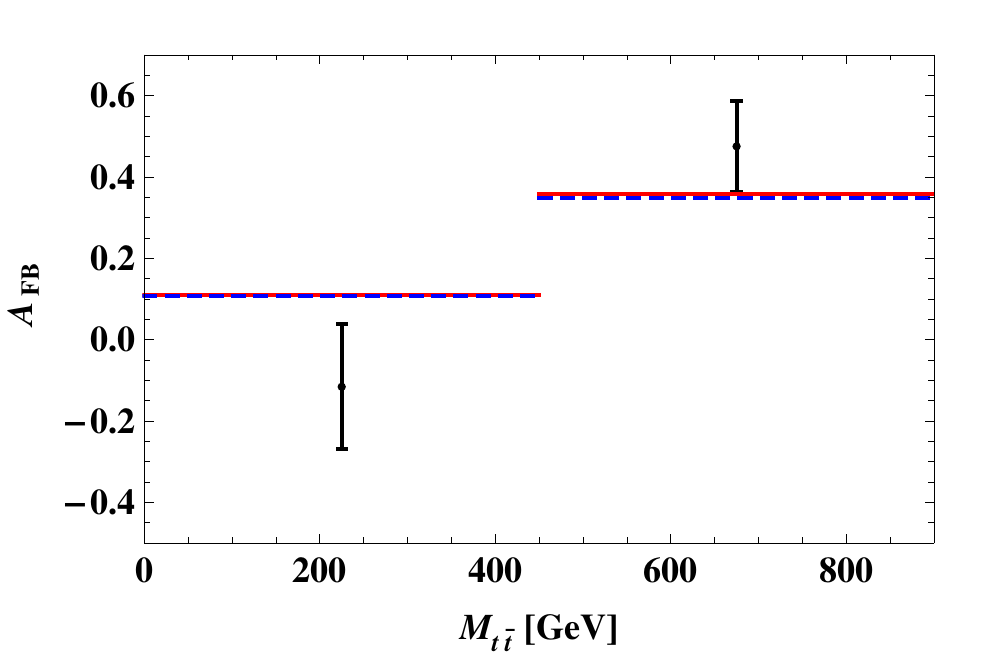}
\includegraphics[width=0.47\textwidth]{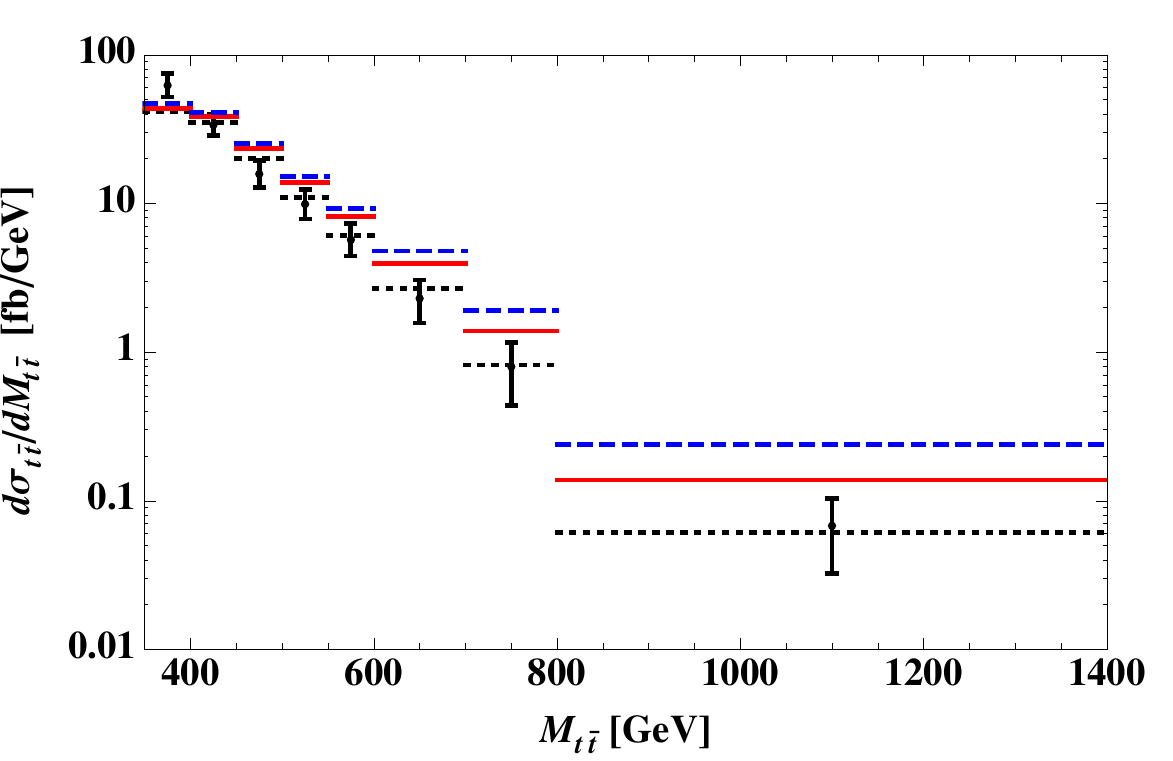}
\caption{$\rm A_{FB}^{t \bar{t}}$ and $d\sigma(t \, \bar{t})/dM_{t\bar t}$ with octet vector exchange, for two 
different values of ($m_{V},  \sqrt{\eta_{ij}\eta_{33}}, \eta_{i3}, \Gamma_V/m_V$): solid red ($300 \, {\rm GeV}, 1,1.33, 0.08$);
dashed blue ($1200 \, {\rm GeV}, 2.2, 4.88, 0.5$), that give approximately the same $\rm A_{FB}^{t \bar{t}}$ in the high mass bin. Taken from \cite{Grinstein:2011yv}. The cross sections still need to be multiplied by the acceptance corrections in Fig. \ref{Fig-eff}.
}\label{Fig.Vector.AFB}
\end{figure}

Beside small FCNCs, fields in flavour nontrivial representations can have other phenomenologically beneficial properties that can help understand a large $\AFBtt$. For instance such fields can have $O(1)$ couplings to quarks even for intergenerational transitions. A flavour octet vector also has the right sign change for a heavy $s$-channel resonance (suppressing colour indices)
\beq
(\bar U_R T^A\gamma^\mu U_R)V_\mu^A=\tfrac{1}{\sqrt3}V_\mu^8(\bar u_R\gamma^\mu u_R+\bar c_R \gamma^\mu c_R- 2 \bar t_R\gamma^\mu t_R)+\cdots,
\eeq
where the sign flip is a consequence of the flavour conserving nature of the interactions and not due to flavour breaking. Similarly, large intergenerational couplings are also flavour conserving since the vectors carry nonzero flavour numbers. For instance for the flavour octet contribution in the $t$-channel we have
\beq
(\bar U_R T^A\gamma^\mu U_R)V_\mu^A=(V_\mu^4-i V_\mu^5)(\bar t_R\gamma^\mu u_R)+\cdots,
\eeq
so that there is a $\mathcal O(1)$ quark generation changing coupling without the need for flavour violation. In $t\bar t$ production the vector then contributes both in the $s$ and $t$ channels, the relative size also depending on the amount of flavour breaking. The predicted $\AFBtt$ and $d\xtt/d m_{t\bar t}$ for two choices of masses and couplings are shown on Fig.~\ref{Fig.Vector.AFB}.

Furthermore, if the NP fields are in nontrivial flavour representations, one avoids pair production of like-sign top pairs from $t$-channel processes (up to small flavour breaking terms). On the other hand, since the fields couple with $\mathcal O(1)$ couplings to all generations, di-jet constraints  are potentially important. Although the predicted di-jet cross sections can be made small enough to avoid present bounds~\cite{Grinstein:2011yv},  this sometimes requires some degree of flavour breaking. The couplings to first generation quarks can still be $\mathcal O(1)$ but smaller than the couplings to the top~\cite{Ligeti:2011vt}. 

An interesting question is whether a single set of fields can explain the $\AFBtt$ as well as the $B_s$ mixing anomaly (c.f.~\cite{Ligeti:2010ia,Lenz:2010gu}) at the same time. If the NP fields couple only to $U_R$ the answer is no. However, if NP fields couple to $Q_L-Q_L$ or $D_R-Q_L$ there are tree level contributions to $B_s$ mixing (if they couple to  $U_R-Q_L$,  $U_R-D_R$ the contributions to $B_s$ mixing arise at the 1-loop level and can also be relevant). The detailed answer depends on the flavour breaking pattern one assumes. In~\cite{Grinstein:2011to-appear} this question is addressed assuming MFV and a number of potentially viable charge assignments for NP fields are identified. 

\subsection{Incoherent production}
A possibility that $\AFBtt$ is due to incoherent production of $t\bar t$+invisible was raised in \cite{Isidori:2011dp}. Unlike the models we discussed so far the large $\AFBtt$ does not arise from  interference of NP amplitude with SM one gluon exchange, but from a large $\AFB$ in the new sector alone (ideally this would be $\sim 100\%$ in the NP cross section). As indicated by the model independent (but only two-bin) analysis of Fig. \ref{f.sigmas} such models cannot give a better than $\sim 2$ sigma agreement with experiments since $\sigma_B^{NP}$ can at best be zero due to lack of interference. Nevertheless, this offers and interesting alternative to other scenarios. 

\begin{figure}[t]
\includegraphics[scale= 1.3]{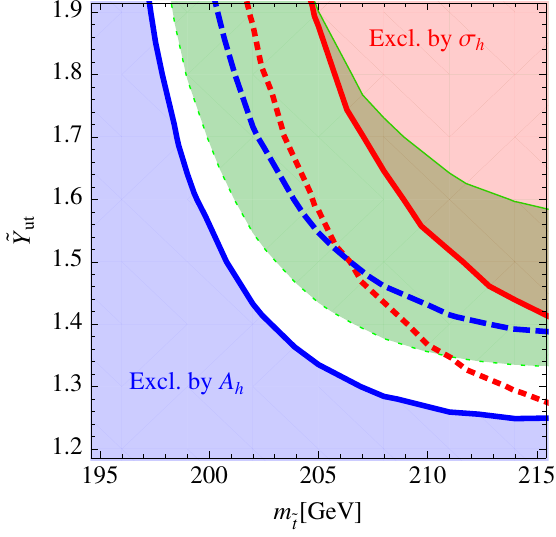}~~~~~~~~
\caption{Tevatron constraints in the $m_{\tilde t}$--$\tilde Y_u$ plane  of incoherent production model~\eqref{eq:Lagr}.
The inclusive $\AFBtt$ and $\sigma_{t\bar t}$ are reproduced within $1~\sigma$ in the 
central green band. The region below the continuous (dashed) blue 
line is excluded by $A_h\equiv A_{FB}(m_{t\bar t}>450 GeV)$ at 95\% C.L. (90\% C.L.). The region above the continuous (dotted) 
red line is excluded by $\sigma_h\equiv \sigma_{t\bar t}(700\,{\rm GeV}<m_{t\bar t}<800\,\rm GeV)$ at 95\% C.L.~(90\% C.L.). Taken from  Ref.~\cite{Isidori:2011dp} .}
\label{fig:parSpace}
\end{figure}

In order to have a large $\AFB$ in the new sector one necessarily needs a $t$-channel contribution from a light state. In \cite{Isidori:2011dp} the authors considered a scenario with a 200 GeV scalar $\tilde t$ with the quantum numbers of the right-handed top, and an $SU(2)_L\times U(1)_Y$ singlet fermion $\chi^0$ with mass of 2 GeV
\bea
\mathcal L &=& \mathcal L_{SM} + (D_\mu \tilde t)^\dagger (D^\mu \tilde t) - m_{\tilde t}^2 \tilde t^\dagger \tilde t +  \bar \chi^0 (i \gamma_\mu D^\mu) \chi^0
 \nonumber\\
&&- m_\chi {\bar\chi}_c^{0}  \chi^0~
+ \sum_{q=u,c,t} (\tilde Y_{q} \bar q_R \tilde t \chi^0 + {\rm h.c.} )~.
\label{eq:Lagr}
\eea
The production process is $p\bar p\to \tilde t\tilde t^\dagger\to t\bar t\chi^0\chi^0$ and a sizable $\AFBtt$ is generated though the exchange of $\chi^0$ in the $t$-channel. Extra missing transverse energy changes the $t\bar t$ spectrum and could be used for detection, once the experiments have enough sensitivity. The allowed parameter space is show on Fig. \ref{fig:parSpace}. $\chi^0$ can be a dark matter candidate, however it cannot be a simple thermal relic~\cite{Isidori:2011dp,Hektor:2011ms}.

%% file: 4s-channel.tex
\section{Predominantly $s$-channel new physics}
\label{sec:schannel}

\subsection{General considerations}

Let us now assume that the NP amplitude is due to an $s$-channel process.  To obtain a nonzero $\AFBtt$ from the interference with the SM $t\bar t$ production amplitude, several requirements on the couplings of the NP resonance to quarks need to be fulfilled. The SM amplitude is $C$ and $P$ even. The NP amplitude can be written as a product of initial and final state vertex contributions.  Applying $C$ only on the final state vertex this interchanges $t$ and $\bar t$ along with changes to the form of the interaction vertex due to the action of the $C$ operator. It then immediately follows that $\AFB$ vanishes, if the interaction is $C$ even. The same argument holds for the initial state interaction. Thus both couplings of the NP resonance to light quarks (initial state) and top quarks (final state) need to be $C$ odd or have a $C$ odd component (and assuming $CP$ thus also parity violating).
 
Lorentz invariance dictates that new particles in the $s$-channel have to have integer spins. An attractive choice is a massive vector boson present in many models beyond SM. In this case, a color octet particle is preferred in order to use its large QCD interference with the one gluon exchange in the SM. Since the QCD interactions already provide the vector couplings for the interference term, only nonzero axial couplings are needed for a spin one, color octet particle $G'$ to produce a nonzero $\AFBtt$. Alternatively, for a color singlet $Z'$, the interference term with the SM is negligible. One then has to have both vector and axial couplings of $Z'$ to $q \bar{q}$ and $t \bar{t}$.  Models of this type can be found in Refs.~\cite{Frampton:2009rk,Chivukula:2010fk,Bai:2011ed,Ferrario:2009bz,Xiao:2010ph,Ferrario:2009bz,Ferrario:2009ee,Martynov:2010ed,Bauer:2010iq,Chen:2010hm,Burdman:2010gr,Degrande:2010kt,Choudhury:2010cd,Cao:2010nw,Foot:2011xu}. 

\subsection{Colour octet vector bosons}
\label{sec:ColorOctets}

We first consider a color-octet resonance $G^\prime$ allowing for the most general renormalizable interactions to quarks, the vector couplings $g_V^{q,t}$ and axial-vector couplings $g_A^{q,t}$. The spin averaged matrix squared for $q\bar q\to t\bar t$ scattering is 
\bea
\sum \left |{\cal M} \right |^2 & = &
g^4_{s}(1+c^2+4m^2)  + \frac{2 g^{2}_{s}\hat{s} (\hat{s}-M_G^2)}
{(\hat{s}-M_G^2)^2+M_G^2 \Gamma_G^2}  \nonumber\\
&&\hspace{-1.9cm}
\times \left[ g_V^q \, g_V^t \, (1+c^2+4m^2) + 2 \, g_A^q \, g_A^t \, c  \right]  \nn \\ 
&&\hspace{-1.9cm}+\frac{\hat{s}^2} {(\hat{s}-M_G^2)^2+M_G^2 \Gamma_G^2}
\left[  (g_V^q)^2+(g_A^q)^2 + 8 \, g_V^q \, g_A^q \, g_V^t \, g_A^t \, c \,  \right.   \nonumber\\
&&\hspace{-1.9cm} \left.\times \left( (g_V^t)^2 (1+c^2+4m^2) +  (g_A^t)^2 (1+c^2-4m^2) \right) \right] \,,
\label{eq:qqtt}
\eea
where $m=m_t/\sqrt{\hat{s}}$, $\beta = \sqrt{1-4m^2}$ is the velocity of the top quark
 and $c = \beta \cos\theta$, with $\theta$ the polar angle of the 
top quark with respect to the incoming up quark in the parton center-of-mass frame.  The color-octet vector resonances are naturally broad $\Gamma_G/m_G \gtrsim O(\alpha_S)$. The terms in Eq. (\ref{eq:qqtt}) that are odd in $c$ generate a charge asymmetry. A positive asymmetry can be generated from the interference term for $g_A^q g_A^t<0$ (assuming $M_G^2>\hat s$), or if the NP squared term $g_V^q g_A^t g_V^q g_A^t c$ dominates.

\begin{figure*}
\begin{center}
\includegraphics[width=.32\textwidth]{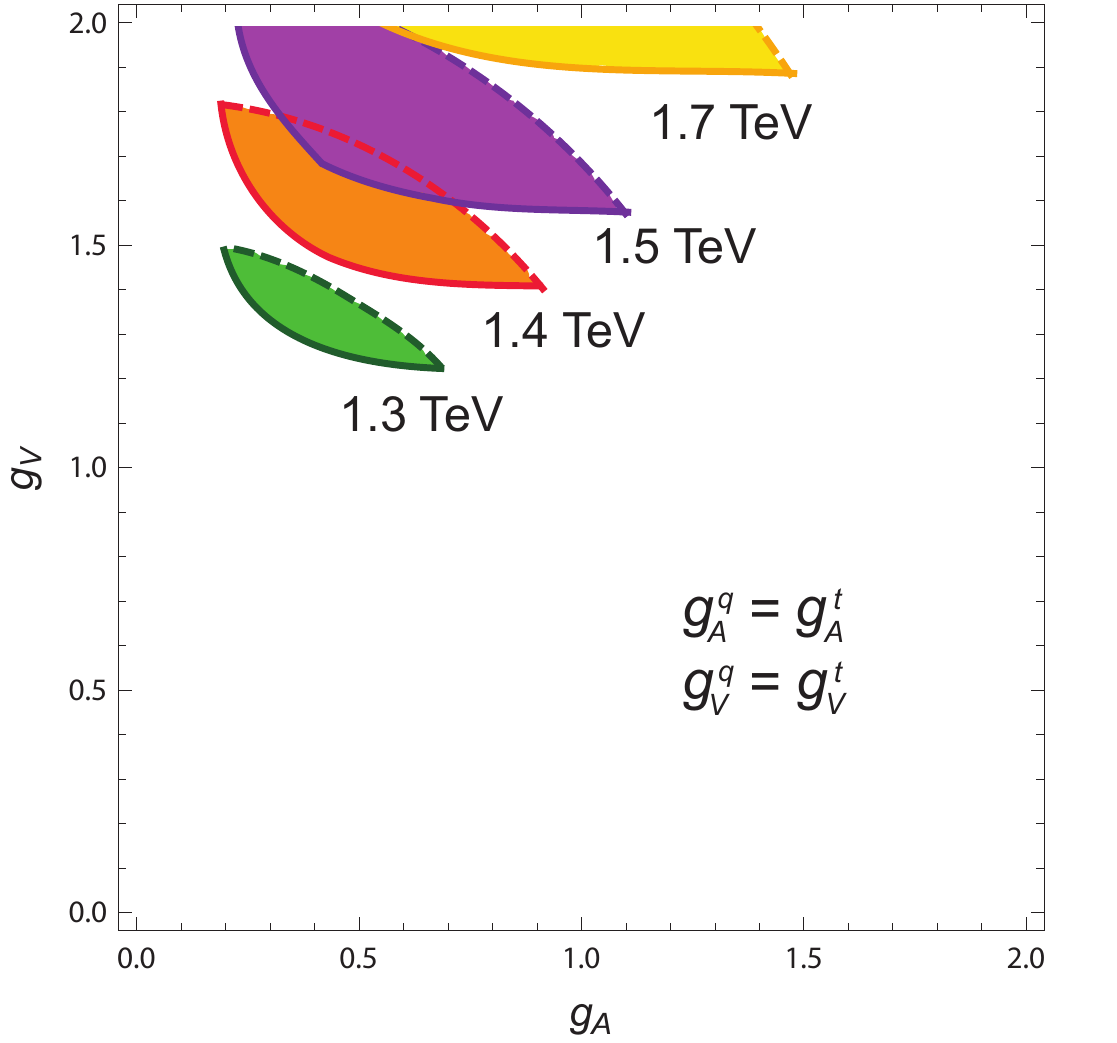} 
\includegraphics[width=.32\textwidth]{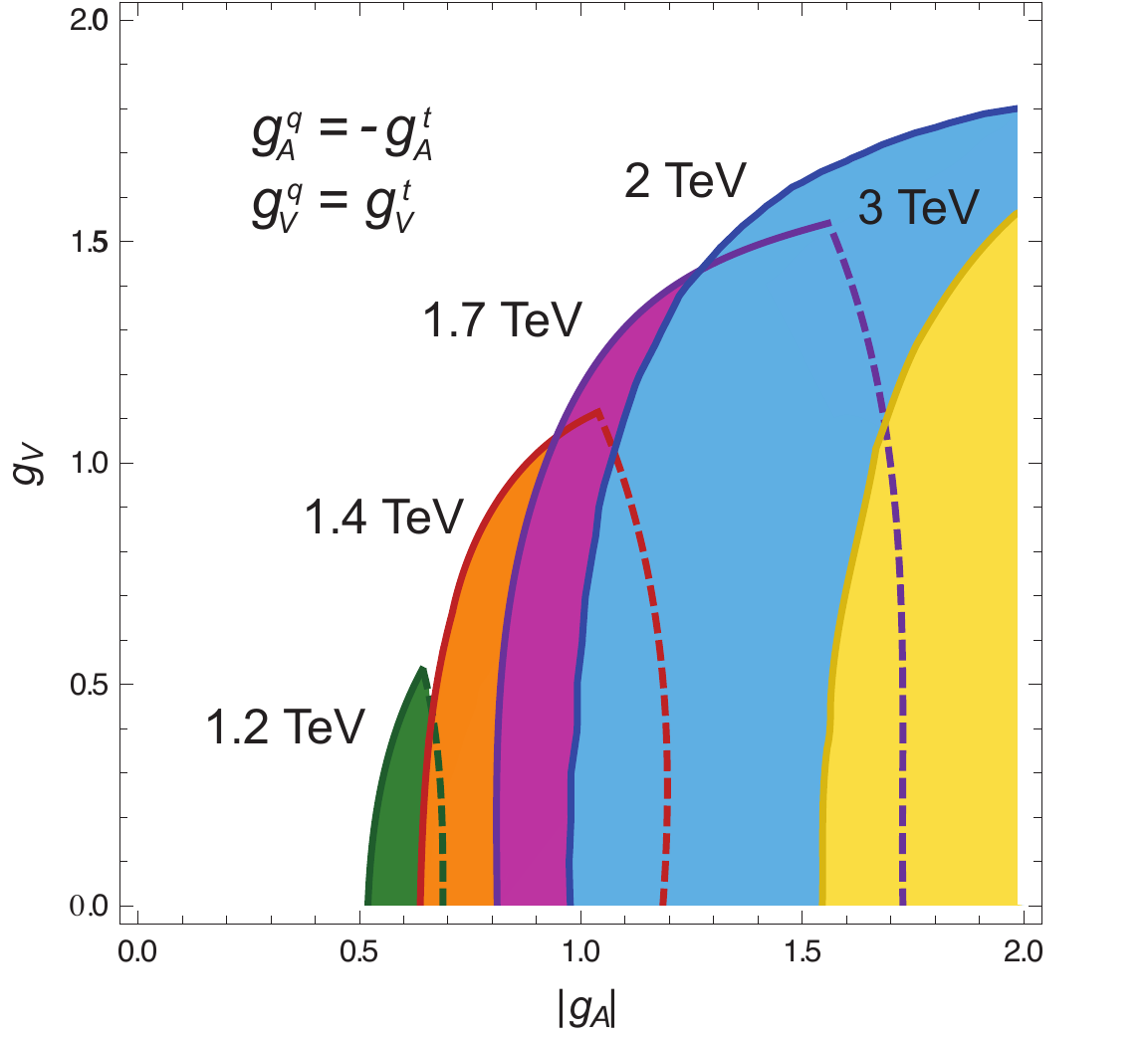} 
\includegraphics[width=.32\textwidth]{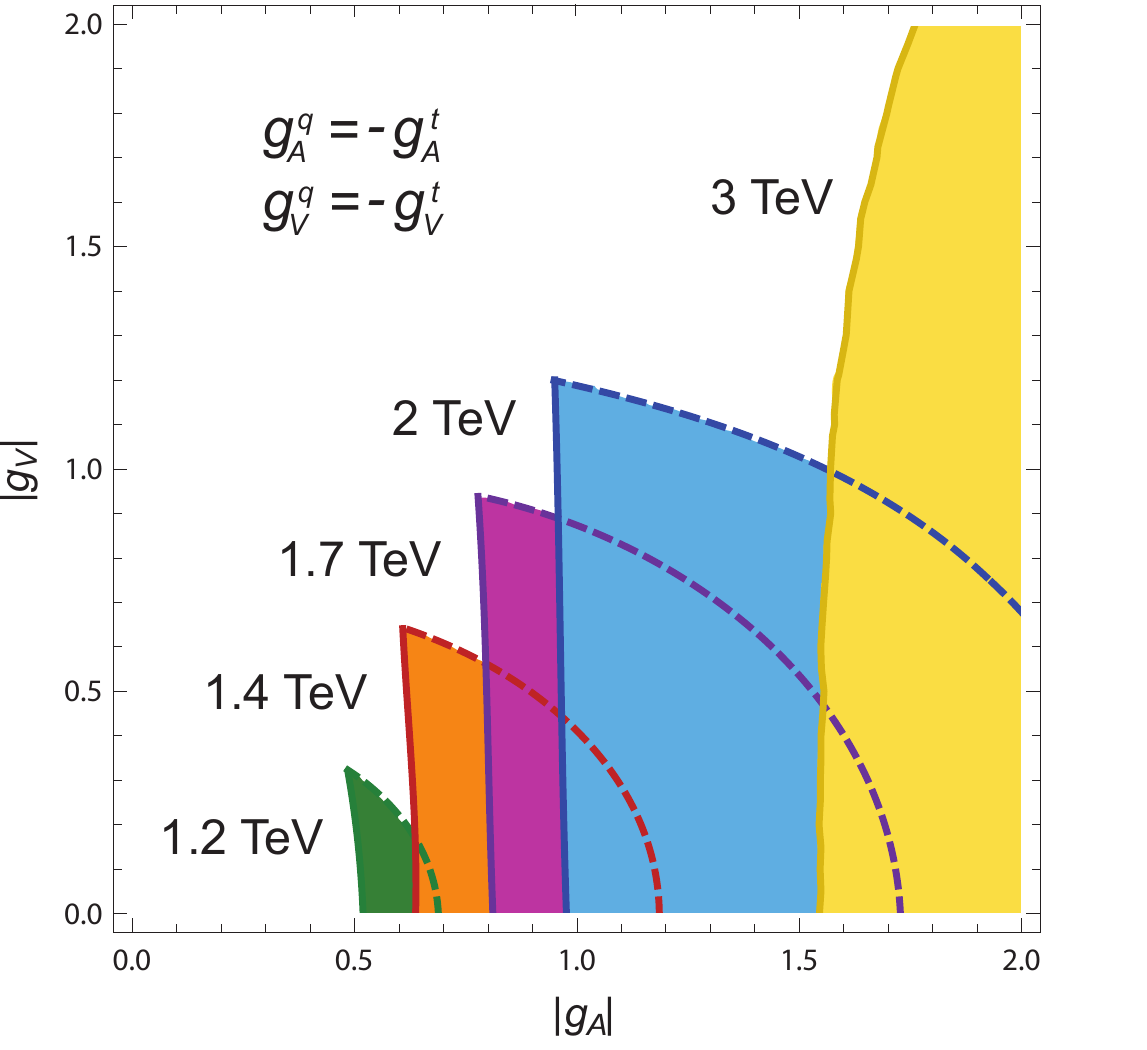} 
\caption{\label{fig:universal} Contours as a function of the vector and axial-vector couplings for different values of the resonance mass for flavor universal couplings 
(left plot: at 95\% C.L. for model A), and flavor non-universal couplings (middle plot, at 90\% C.L. for model B) (right plot, at 90\% C.L. for model C). Taken from Ref. \cite{Ferrario:2009bz}. }
\end{center}
\end{figure*} 
 
In Fig.~\ref{fig:universal} we quote constraints on such ``axigluon" models obtained in Ref. \cite{Ferrario:2009bz} from the inclusive $\AFBtt$~\cite{oldcdf} 
and $d\xtt/dm_{t\bar t}$~\cite{Aaltonen:2009iz} measurements. They considered three different types of axigluon models 
\begin{itemize}
\item[] {{\bf Model A}: flavor universal axigluon, $g_V^q=g_V^t$ and $g_A^q=g_A^t$.
}
\item[] {{\bf Model B}: flavor nonuniversal axigluon with axial-vector couplings to light quarks and top quarks of opposite signs, $g_V^q=g_V^t$ and $g_A^q=- g_A^t$.}
\item[] {{\bf Model C}: flavor nonuniversal axigluon with opposite signs in both vector and axial-vector couplings: $g_V^q=-g_V^t$ and $g_A^q=- g_A^t$.}
\end{itemize}
We consider $M_{G'}\gtrsim 1$ TeV. If the axigluon has flavor universal couplings, these are well constrained due to the negative asymmetry in the interference term. Much larger parameter space is allowed for the flavor nonuniversal models since these lead to large positive asymmetry in the interference term. Note that in Ref.~\cite{Ferrario:2009bz} only bounds from the the inclusive $\AFBtt$~\cite{oldcdf} and the differential cross section~\cite{Aaltonen:2009iz} measurements were considered.

\begin{figure*}[t]
\begin{center}
\vspace*{3mm}
\includegraphics[width=0.35\textwidth]{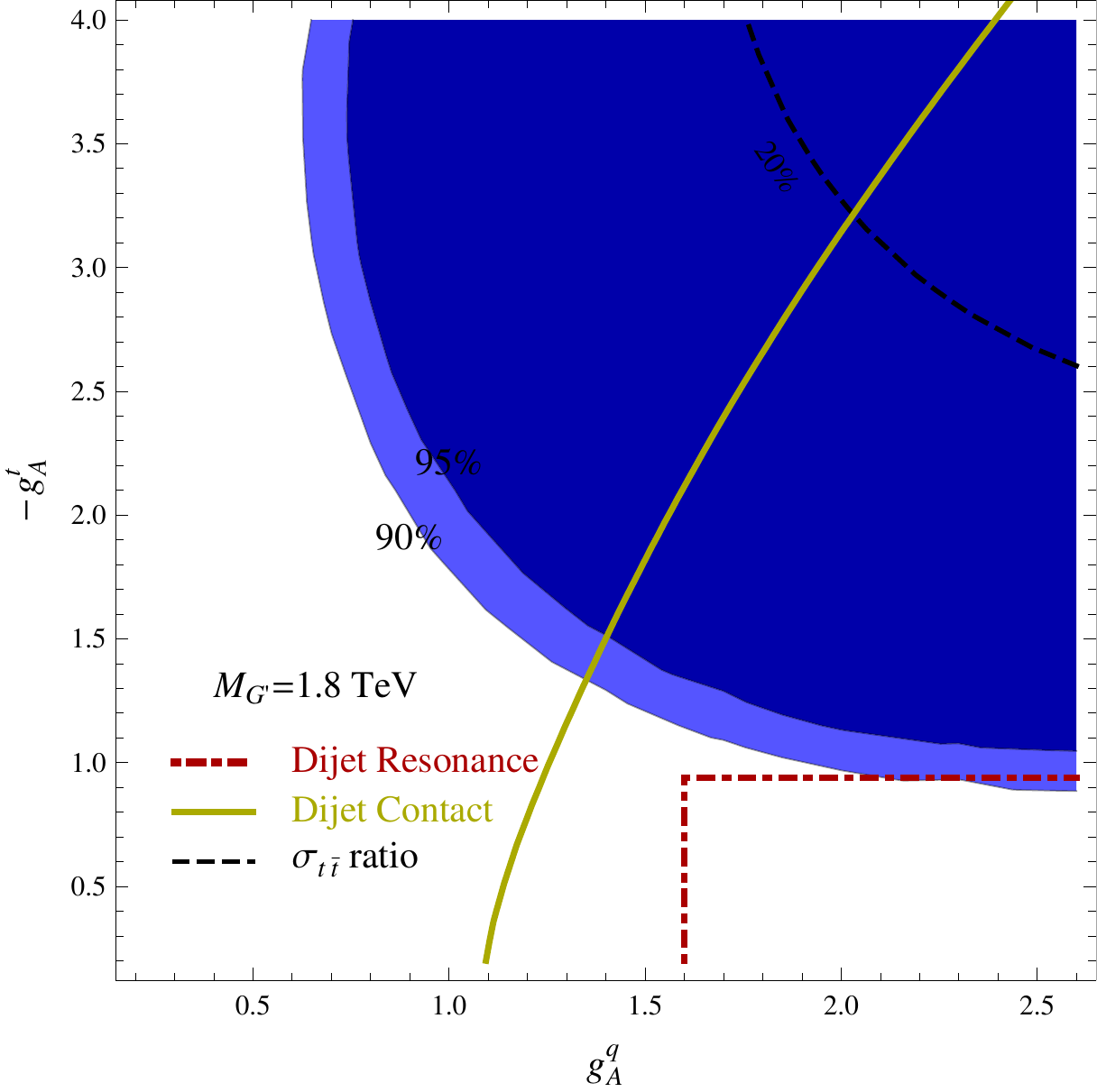} \hspace{1cm}
\includegraphics[width=0.35\textwidth]{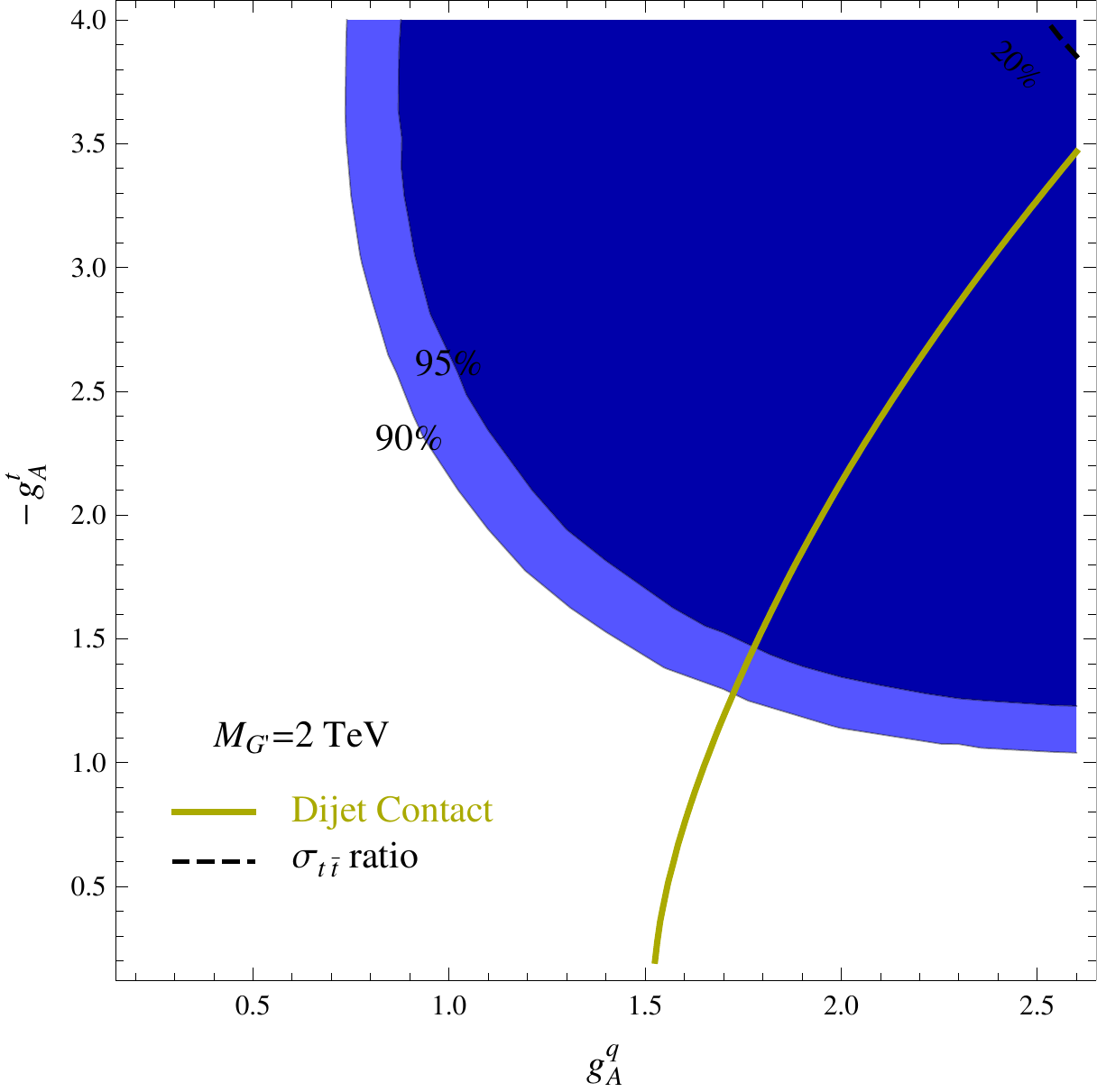} 
\caption{The constraints on axigluon with $g_A^q > 0$, $g_A^t < 0$ and $g_V^{q, t}=0$ and mass
$M_{G^\prime} =1.8$~TeV (left) or $M_{G^\prime} =2$~TeV (right). The region on the right side of the yellow solid line is excluded by dijet contact interaction searches. The narrow resonance dijet searches do not further constrain the parameter space. Taken from Ref. \cite{Bai:2011ed}. }
\label{fig:generalmodel}
\end{center}
\end{figure*}

More recent constraints are shown on Fig.~\ref{fig:generalmodel}. Taking into account the CDF measurements of $\AFBtt$ for the different $t \bar{t}$ invariant mass and rapidity bins listed in Table~\ref{table:partondata}, the total cross section 
(but not the $m_{t\bar t}$ distribution) and constraints from the early di-jet resonance~\cite{Khachatryan:2010jd} and contact interactions~\cite{Collaboration:2010eza} searches at the LHC, Ref.~\cite{Bai:2011ed} derived limits on the purely axially coupled $G'$ ($g_V^{q,t}=0$) with $g_{A}^{q}g_{A}^{t}<0$. The LHC di-jet measurements in particular, severely constrain axigluon like models, and the most recent results~\cite{dijetEPSATLAS,dijetEPSCMS} already completely rule out the simplest scenarios in which one assumes $g^q \sim -g^t$. For instance, the sample point $M_{G'}= 2$ TeV, $g_A^q=$ 2.2, $g_A^t=$-3.2 and $g_V^q = g_V^q =0$ considered in Ref.~\cite{Bai:2011ed} is already ruled out. Conversely, only models with $|g^q| \ll |g^t|$ are allowed.

Another interesting possibility is that the color-octet vector boson is purely axial and light, with mass below or close to the $m_{t\bar t}=450$~GeV edge separating the experimental $\Alow$ and $\Ahigh$ bins~\cite{Xiao:2010ph,Tavares:2011zg}. The deviation in the $t \bar{t}$ differential cross section is smaller because the interference with the SM cancels in the cross section (but not in the $\AFBtt$). Another nice feature is that the $\AFBtt$ automatically changes sign when $m_{t \bar{t}}$ passes the light resonance mass. To obtain a large enough $\AFBtt$, already relatively weakly coupled axigluons with $g_A\sim g_s/3$ suffice. Constraints from $d\xtt/dm_{t\bar t}$ and di-jets are still important, but can be avoided even for flavor universal couplings $g_A^q=g_A^t$ if the axigluon has a large enough decay width $\Gamma_G \sim 0.2 M_G$~\cite{Tavares:2011zg}.

The phenomenological description of a single axigluon coupling to quarks requires a UV completion. The reason is that the for a nonzero axial-vector coupling between $G'$ and fermion $\psi$ (e.g. top quark), the amplitude for the process $\bar{\psi} \psi \rightarrow G' G'$ grows as \cite{Shu:2007wg}
\bea
\mathcal{M} \sim 4 (g_A^\psi)^2 \frac{m_\psi}{M_{G}} \sqrt{\hat s}\,,
\eea
for $\hat s \gg m_\psi^2, ~ M_G^2$. Therefore there is a tree level unitarity bound 
\bea
\sqrt{\hat s} \lesssim \Lambda_U = \frac{ \sqrt{2} \pi M_{G}^2}{C_F (g_A^\psi)^2 m_\psi }\,,
\eea
with $C_F = 4/3$. In order to explain the large $\AFBtt$ one has $\Lambda_U\sim {\mathcal O}(30)$ TeV taking $g_A^q\sim g_A^t$. It is smaller, if coupling to top quarks is larger (for instance as may be required by di-jet constraints).

\subsection{Chiral colour models}
Colour-octet gauge bosons appear in gauge extensions of the SM which are spontaneously broken to $SU(3)_c$. The simplest possibility is $SU(3)_A \times SU(3)_B\to SU(3)_c$~\cite{Frampton:2009rk}. In order to have a large positive asymmetry in the interference term one needs $g_A^q g_A^t < 0$ (assuming $M_G$ is well above $500$ GeV). An assignment of gauge quantum numbers giving this is that $t_R$, $b_R$ are triplets of $SU(3)_B$ while $(t, b)_L$, $q_R$ are triplets of $SU(3)_A$. In order to cancel gauge anomalies a fourth generation charged in the same way as the third is needed. 
The full field content of the model is summarized in Table~\ref{table:assign}. The bi-triplet scalar field $\Sigma$ has a TeV scale vacuum expectation value $\langle \Sigma_{i\bar{k}} \rangle = u \delta_{i\bar{k}}$ breaking $SU(3)_A\times SU(3)_B$ to the QCD color group $SU(3)_c$. There are two Higgs scalars giving masses to quarks, the quarkonic Higgs $H_q$ which is a triplet of  $SU(3)_A$ and $SU(3)_B$, and the leptonic Higgs $H_l$ which is a singlet.  In Ref.~\cite{Chivukula:2010fk} it was pointed out that such a model generically suffers from too large FCNC contributions to $B_d-\bar{B}_d$ mixing. It is however possible to protect it by flavor symmetries so that all contributions to meson mixings are in agreement with experimental constraints~\cite{Shu:2011au,Bai:2011ed}.
\begin{table*}
\begin{center}
{\renewcommand{\arraystretch}{1.1}
\begin{tabular}{c|ccc|ccc|cc|ccc}
  \hline\hline
  \rule[5mm]{0mm}{0pt}
Field & $Q_i$ & $u^{c}_i$ & $d^{c}_i$ & $Q_j$ & 
$u^{c}_j$ & $d^{c}_j$ &  $\Sigma$ & $H_{q}$ & $L_k$ & $e^c_k$ & $H_{l}$ \\
\hline
\rule[5mm]{0mm}{0pt}
SU(3)$_A$ &  \bf{3} & \bf{1} & \bf{1} & \bf{1} & 
$\bar{\bf{3}}$ &  $\bar{\bf{3}}$ &  \bf{3} &  \bf{3} & \bf{1} & \bf{1} & \bf{1}\\
 \rule[5mm]{0mm}{0pt}
SU(3)$_B$ &  \bf{1} & $\bf{\bar{3}}$ & $\bf{\bar{3}}$ & 
\bf{3} &  \bf{1} &  \bf{1} &  $\bf{\bar{3}}$  & 
$\bf{\bar{3}}$ & \bf{1} & \bf{1} & \bf{1} \\
 \rule[5mm]{0mm}{0pt}
SU(2)$_{L}$ &  \bf{2} &  \bf{1} &  \bf{1} & \bf{2} &  
\bf{1} &  \bf{1} &  \bf{1} &  \bf{2} & \bf{2} & \bf{1} & \bf{2} \\
 \rule[5mm]{0mm}{0pt}
U(1)$_{Y}$ & ${1/ 3}$ & $-{4/ 3}$ & ${2/3}$ &  
${1/3}$ & $-{4/3}$ & ${2/3}$ & 0 & 1 & -1 & 2 & 1\\
\hline\hline
\end{tabular}
}
\caption{Charge assignment of all the quark, lepton
and Higgs fields under $SU(3)_A\times SU(3)_B\times 
SU(2)_L\times U(1)_Y$. The flavor indices are 
$i=1,2$, $j=3,4$ and $k=1,2,3,4$. The normalization of hypercharge is 
such that the electric charge is given by
$q= I^3_L+Y/2$ with $I^3_L$ the third component 
of $SU(2)_L$ isospin. All the fermion fields are left-handed. Taken from Ref.~\cite{Frampton:2009rk}.}
\label{table:assign}
\end{center}
\end{table*}

The kinetic term for the link field becomes the 
mass term for the massive gauge boson 
$
\rm{Tr}[(D_{\mu} \Sigma)^{\dag} (D_{\mu} \Sigma)] 
\supset {u^2} ( g_A A_{\mu} - g_B B_{\mu})^2 /2
 = {u^2 g^2} (G_{\mu}^{1})^2 /2 
$, where $g \equiv \sqrt{g_A^2 + g_B^2} $. 
The rotation matrix between gauge bosons in mass eigenstates
and gauge eigenstates is
\begin{eqnarray}
\begin{pmatrix}
G^1_{\mu} \\ G^0_{\mu}
\end{pmatrix}
=
\begin{pmatrix}
s_g&-c_g\\
c_g&s_g
\end{pmatrix} 
\begin{pmatrix}
A_{\mu} \\ B_{\mu}
\end{pmatrix}
\ ,
 \label{G_MG}
\end{eqnarray}
where we define $s_g \equiv \sin\theta \equiv  g_A/g$ and 
$c_g \equiv  \cos \theta = g_B/g$ so 
$\theta = \arctan (g_A/g_B)$. The massless 
field $G^0_\mu$ is the usual QCD gluon while we identify 
the massive octet vector boson $G^1_\mu$ with the axigluon.  

\begin{figure}[t]
\begin{center}
\includegraphics[angle=0,width=8cm]{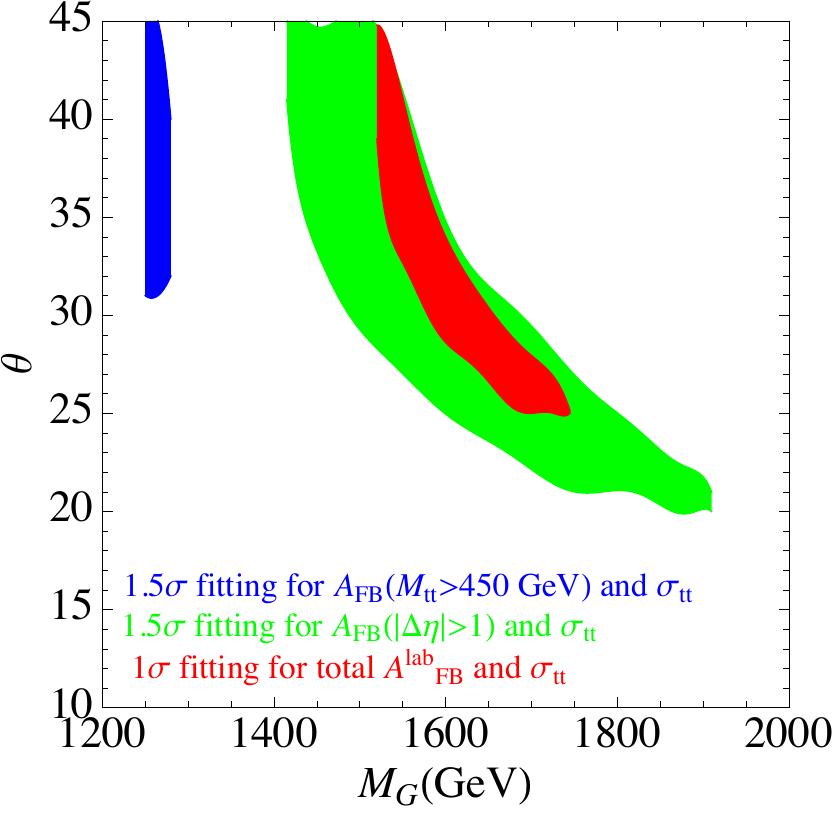}
\caption{1~$\sigma$ parameter region coming from the $\AFB^{\rm inc.}$ in the $p\,\bar p$ frame ($\AFB^{\rm lab}$) and the inclusive $t\bar{t}$ production cross section ($\sigma_{t\bar t}$)
(shaded in red). Also shown are the 1.5~$\sigma$ parameter regions from $\AFB(M_{t\bar{t}}>450)\equiv\Ahigh$, $\sigma_{t\bar t}$ (in blue), and from $\AFB(|\Delta\eta|>1)\equiv \AFBtt(|\Delta y| > 1)$, $\sigma_{t\bar t}$ (in green).  Taken from Ref.~\cite{Shu:2011au}.
}
\label{fig:axigluon}
\end{center}
\end{figure}

The QCD couplings are flavor universal with $g_s = g s_g c_g$. There is a $g_A \leftrightarrow g_B$ symmetry in the model of $\theta \leftrightarrow 90\degree - \theta$. Requiring that the model be perturbative, $g_A, g_B < 2 \pi$, imposes
 $10\degree < \theta < 45 \degree$.
Fermions charged under $SU(3)_A$ 
and $SU(3)_B$ couple to the massive 
axigluon $G^1_\mu$ with $g s_g^2$ and $-g c_g^2$, 
respectively. Therefore, the vector and axi-vector coupling strengths are   
\beq
g^{q}_{V} = g^{t}_{V} = - \frac{g\, {c_{2g}}}{2},~~- g^{q}_{A} =  g^{t}_{A} = \frac{g}{2}\,,
\eeq 
where $c_{2g} \equiv \cos (2 \theta)$. The axial-vector couplings are thus always nonzero and satisfy the following relations
\bea
\label{eq: condition}
g_A^q g_A^t <0, ~~g_V^q g_V^t >0~~~  \mathrm{with }~~~ g_V < g_A \ ,
\eea 
which are crucial for successful phenomenology (see Sec.~\ref{sec:ColorOctets}). The preferred regions to explain a large $\AFBtt$ are shown in 
Fig. \ref{fig:axigluon}.

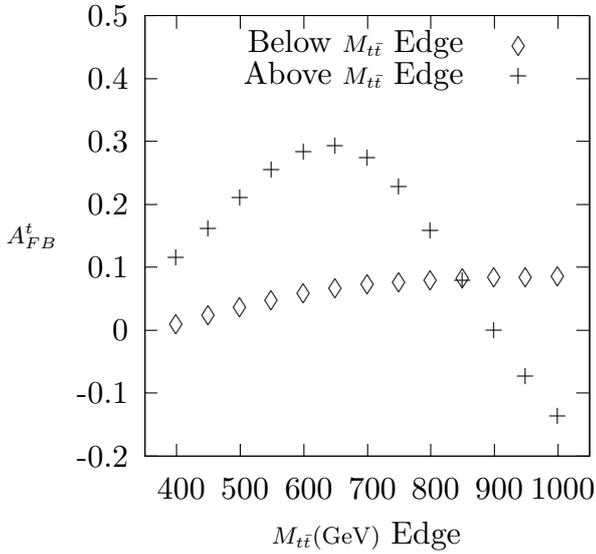
\begin{figure}[t]
\input{afb_mtt2.tex}
\caption{$\AFB^t\equiv \AFBtt$ above/below a $M_{t\bar{t}}\equiv m_{t\bar t}$ edge using the benchmark point, $M_{G}=1525~$GeV, 
 $g^q_A=-g^t_A=-1.155g_s$ and $g^{q}_{V}=g^{t}_{V}=-0.577g_s$. Taken from Ref.~\cite{Frampton:2009rk}.
}
\label{Fig:afbmtt2}
\end{figure}

An interesting feature of  the model~\cite{Frampton:2009rk} is the mass dependent
asymmetry. Due to the opposite contributions to the
asymmetry from the interference term and the
NP squared terms, the asymmetry is positive
at intermediate $m_{t\bar t}$ but becomes negative
close to the $t\bar t$ threshold. This change of 
 $\AFBtt$ sign with  $m_{t\bar{t}}$ has been observed by CDF both in measurements with several $m_{t\bar{t}}$ bins
as well as in the simple two-bin measurement of $\Alow$ and $\Ahigh$ (see Fig. \ref{Fig:afbmtt2}).It has been shown recently~\cite{Tavares:2011zg}, that the same effect can be achieved by using a light resonance with a purely axial-coupling. For $\AFBtt$ dominated by the interference term  the sign change occurs when $m_{t \bar{t}}$ passes the resonance mass. If one chooses $g_a \sim g_s /3$, one can predict a negative $\Alow$.

The simplest model of Ref.~\cite{Frampton:2009rk} is ruled out by the recent LHC di-jet measurements~\cite{Khachatryan:2010jd,Collaboration:2010eza,dijetEPSATLAS,dijetEPSCMS}. To avoid them one needs to tune the relative ratios between the $G'$ top couplings and the $G'$ light quark couplings to suppress the $G'$ decay branching ratio to di-jets. One simple solution is to introduce a vector-like fermion mixing with the SM light fermions. (Notice that this is the way to deconstruct~\cite{ArkaniHamed:2001ca} a chiral fermion in the two site model). For simplicity, we only introduce one vector-like fermion which mixes solely with the up quark.
The fermions $\psi_{L,R}$, are $SU(2)_W$ singlets and have charge $2/3$ under $U(1)_Y$. Under the extended gauge group  
$q_L$, $q_R$, $t_R$, $b_R$, $\psi_L$ are assigned to be triplets of $SU(3)_B$ and $(t, b)_L$, $\psi_R$ to be triplets of $SU(3)_A$. In the up-type quark sector, the general $4\times 4$ mass matrix can be diagonalized by a bi-unitary transformation acting on the left-handed and right-handed quarks. 
The transition from the flavor basis to the mass eigenstate is then parametrized by
\bea
u^{(m)}_R &=&  \cos{\alpha} \,  u^{(f)}_R  \,+\, \sin{\alpha}\, \psi^{(f)}_R\,, \nonumber \\
\psi^{(m)}_R &=&  -\sin{\alpha} \,  u^{(f)}_R   \,+\, \cos{\alpha}\, \psi^{(f)}_R\,. 
\eea
The couplings of the axigluon $G^\prime_\mu$ to the various quarks in the mass eigenstate basis are found to be
\begin{align} 
&g_V^{(d, s, b, c)} = - \tan{\theta} \,,&& g_A^{(d, s, b, c)} = 0 \,, \nonumber\\
&g_V^t = \frac{1}{\tan{2\theta}}\,,&& g_A^t = 
\frac{1}{\sin{2\theta}} \,, \nonumber \\
&g_V^{u} =  - \tan{\theta} \,+\, \frac{\sin^2{\alpha}}{\sin{2\theta}} \,,&&
     g_A^{u} = -  \frac{\sin^2{\alpha}}{\sin{2\theta}} \,.
 \label{eq:vectorfermioncoupling}
\end{align}
When $\alpha=\pi/2$, the couplings of the up and top quarks become identical to those in the minimal two-site model. Notice that the axial-vector couplings of the up and top quarks are different and one now has the freedom to increase $g_A^t$ by reducing $\theta$ 
and simultaneously decrease $g_{V, A}^u$ by choosing $\sin^2{\alpha}$ smaller than $\sin{2\theta}$. The total width of the 
$G^\prime_\mu$ in this model is
\bea
\Gamma(G^\prime) &=& \frac{\alpha_s\,M_{G^\prime}}{12}\,\left[(\cos^4{\alpha}+10) \tan^2{\theta} \right. \nonumber\\
&&\left. + (\sin^4\alpha + 1) \cot^2{\theta} - 2 
\sin^2\alpha \cos^2 \alpha \right]   \,.
\label{eq:width_vector}
\eea

\begin{figure}[t]
\begin{center}
\vspace*{3mm}
\includegraphics[width=0.45\textwidth]{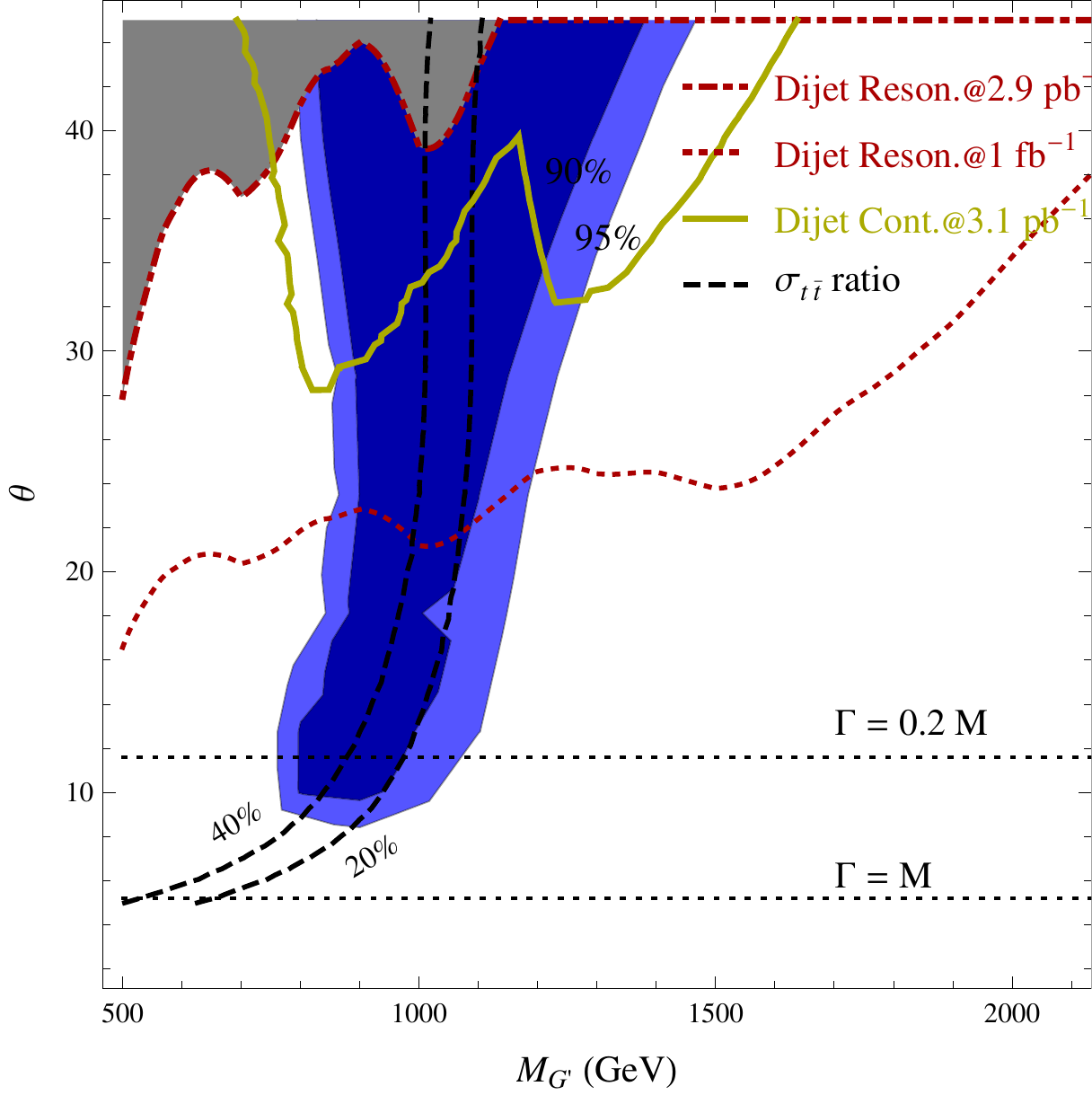} 
\caption{The fit of the model with one axigluon and one additional vector-like fermion (see text) to  $\AFBtt$ observed by CDF (shaded in blue). The region above the red dot-dashed (yellow solid) line is excluded at 95\% C.L. 
by the first CMS di-jet narrow resonance~\cite{Khachatryan:2010jd} (ATLAS contact interactions~\cite{Collaboration:2010eza}) searches. Red dotted line shows the projected exclusion limit from the di-jet narrow resonance search at the $7$ TeV LHC with 1 fb$^{-1}$. 
The black dashed lines denote the regions where the given percentage of $\xtt(m_{t\bar{t}} > 450~\rm GeV)$ arises 
from NP. Taken from Ref.~\cite{Bai:2011ed}.}
\label{fig:twositefermion1}
\end{center}
\end{figure}

When $\sin{\alpha} = \sqrt{2} \sin{\theta}$ for $0 \leq \theta \leq \pi/4$, $g_V^u =0$ and the interference between the QCD gluon and the axigluon vanishes the production cross section. Restricting to this part of parameter space the results of the fit are shown in Fig.~\ref{fig:twositefermion1}. 
From the blue-shaded contours one can see that for a small mixing angle the best-fit region is insensitive to its exact value. 
This can be easily understood since the product of the axial-vector couplings $g^u_A g^t_A = -1/(2\cos^2{\theta})$ is insensitive to $\theta$ for small 
$\theta$. The two horizontal dotted black lines with different width/mass ratios show that in the best-fit region  
$G^\prime$  is a narrow resonance. The red dot-dashed line marks the constraint from the first di-jet narrow resonance search at CMS with $2.9$~pb$^{-1}$~\cite{Khachatryan:2010jd}, while the region above the dark yellow solid line is excluded by the search for di-jet contact interactions from ATLAS with 3.1 pb$^{-1}$~\cite{Collaboration:2010eza}. Finally, red dotted line is the projected exclusion limit for 1 fb$^{-1}$ at the LHC -- a situation which is comparable with the most recent experimental results~\cite{dijetEPSATLAS,dijetEPSCMS}. From this figure we conclude that in the two-site plus one vector-like fermion model there still exists a small region in parameter space which is at present allowed by all constraints.

\subsection{RS like models with strong dynamics}
The di-jet searches at the LHC tightly constrain the $G'$ decay branching ratio to light quarks. Therefore a large coupling between $G'$ and the top quark is preferred in order to obtain a large deviation in $\AFBtt$. This can happen if $G'$ is a gluon excitation of some strongly coupled sector which also strongly couples to the top quark. One good example is the RS model.

The RS model is based on a slice of AdS$_5$ background metric $ ds ^2 = ( {z_h}/{ z} )^2  \left[ \eta_{ \mu \nu } 
d x ^{\mu } d x^{ \nu } + ( d z ) ^2  \right] ,$
with curvature $\kappa  = 1/z_h \lesssim M_{Pl}$. Here $x^\mu$ and $z$ are the coordinates of the
standard four dimensions and the extra dimension respectively ($\eta_{\mu \nu} = Diag(-, +, +, +)$).
The UV boundary is at $z_h = 1/ \kappa$
where the scale factor $(z_h / z)^2 = 1$ and the IR boundary is at
$z_v \sim 1 / {\rm TeV}$, as motivated by the hierarchy problem.
We are particularly interested in a model where all SM fields, except perhaps the Higgs, propagate in the entire 5-d spacetime,
and will be primarily concerned with the gluon and colored fermion
fields. The action for the gauge fields and fermions is,
\bea
S &=& \int d^5x \sqrt{-g} \left\{-\frac{1}{4g_5^2} F_{MN}^{a}F^{MN \, a}
+ i \overline{\Psi} \Gamma^{\dot{M}} e_{\dot{M}}^{M} 
D_M \Psi \right.\nonumber\\
&&+ i c \kappa \overline{\Psi} \Psi 
\bigg\}\,,
\label{eq:action0}
\eea
where $\Gamma^{\dot{M}}$ 
are the 5d ($4 \times 4$) Dirac matrices, $e_{\dot{M}}^{M}$ is the veilbein,
$a$ is an adjoint gauge index and $c$ parameterizes the magnitude
of a bulk mass for the fermion in units of the curvature.

\begin{figure}[htb]
\begin{center}
\includegraphics[width=.47\textwidth]{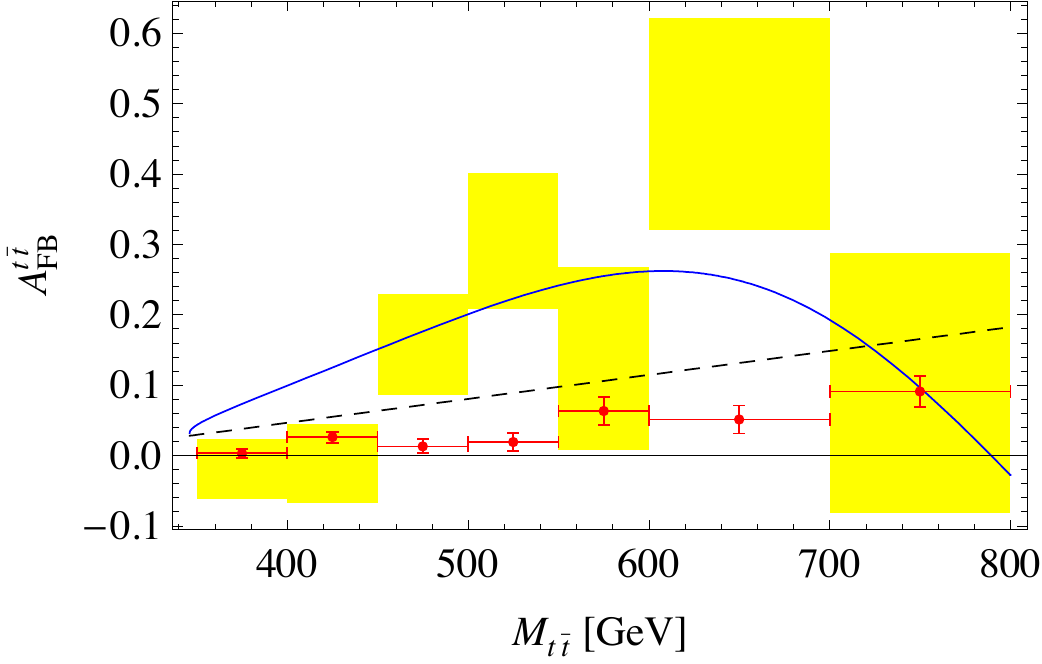}
\caption{The $\AFBtt$ as a function of $M_{t\bar t} \equiv m_{t\bar{t}}\,$.  The RS model \eqref{cset1}  prediction
(including the SM contributions) is in a solid blue line, while the CDF measurement
(at the detector level)~\cite{Aaltonen:2011kc} is described by the yellow
shades. The black dashed line stands for the SM
partonic level prediction, while the red circles with
error bars correspond to the detector level prediction from
MC@NLO~\cite{Aaltonen:2011kc}. From Ref.~\cite{Delaunay:2011vv}. }
\label{fig:afb}
\end{center}
\end{figure}

\begin{figure}[t]
	\centering
	\vspace*{.5cm}
			\includegraphics[width=0.47\textwidth,angle=0]{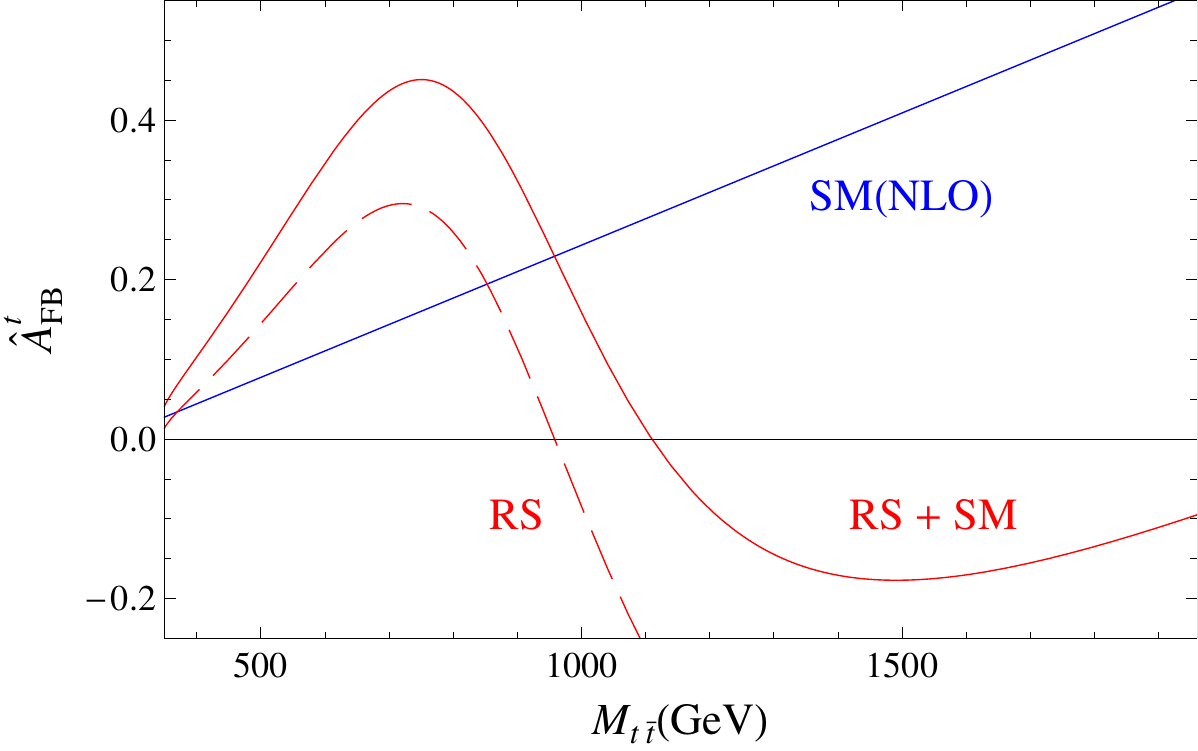}
\caption{\label{partonic} \small{The $\AFB$ for the partonic process $q \bar q \to t \bar t$  ($\hat A_{\rm FB}^t$) as a function of $M_{t\bar t} \equiv m_{t\bar t}$
in the SM at NLO (blue line), at LO with the RS \eqref{cset} contribution (dashed red curve) and at NLO (being the sum of the previous ones) (plain red curve). From Ref.~\cite{Djouadi:2011aj}. }}
\end{figure}

In the RS model the $G'$ which contributes to the $\AFBtt$ is the 1st Kluza-Klein excitation of the gluon. 
In the unitary gauge $A_5 = 0$, we decompose the 5d fields into KK modes,
\bea
A^a_\mu (x,z) = \sum_n A^{a(n)}_\mu(x) g^{(n)}(z) \, , \\
\Psi_{L,R} (x,z) = (\kappa z)^{3/2} \sum_n \psi^{n}_{L,R}(x) 
\xi^{(n)}_{L,R} (z) ~.
\eea
The wave function of the 1st KK gluon is given by 
\bea
g^{(1)}(z) & = & N_1
\: z \: \left[ J_1(m_1 z)+ b_1 Y_1(m_1 z)\right]\,,
\label{eq:KKgluon}
\eea
with normalization factor $N_1$ and
admixture controlled by $b_1$. The mass spectrum is controlled by the
boundary conditions, with the masses satisfying,
\bea
b_1 = - \frac{J_0 \left( m_{1} z_h \right)}{Y_0 \left(  m_{1} z_h  \right) } 
=  - \frac{J_0 \left(  m_{1} z_v \right)}{Y_0 \left( m_{1} z_v \right)} \,,
\label{eq:b0}
\eea
where $m_1$ is related to the first root of the Bessel function. Of particular note 
for the following is the fact that the light KK states
have most of their support close to the IR boundary.
For the fermion zero mode (SM fermion), the wave function is 
\bea
\xi^{(0)}(z, c) & = & N_0 (\kappa z )^{1/2 - c}\,,
\label{eq:KKgluon}
\eea
and the relative coupling ratio between $G'$ and light fermions over the SM strong coupling $g_s$ is the ratio between the integration of 5D the wave-functions
\begin{gather}
g^\psi(c) = 
\frac{\int dz ~ g^{(1)}(z) (\xi^{(0)}(z, c))^2 }
{\int dz  ~ g^{(0)}(z) (\xi^{(0)}(z, c))^2 }~ .
\end{gather}
Therefore, one can adjust the coupling easily by the changing the 5D fermion bulk mass $c$. 
However, localizations of the light fermions are constrained by electroweak precision tests (for a detailed study, see Ref.~\cite{Davoudiasl:2009cd}) and the resulting parameter sweet spots  were found in Ref.~\cite{Djouadi:2011aj}
\begin{align}
c_{u_L}&=c_{d_L} = c_{c_L}=c_{s_L}  \simeq 0.5, & c_{t_L} &= c_{b_L} \simeq 0.02, \nonumber\\
c_{u_R}&=c_{c_R} \simeq 0.15, &c_{t_R} &\simeq 0.48\,,
\nonumber \\
c_{d_R}&=c_{s_R} \simeq 0.63, &c_{b_R}&\simeq0.57\,,
\label{cset}
\end{align}
and Ref.~\cite{Delaunay:2011vv}\,\footnote{Note that this model does not implement an $SU(2)$ flavor symmetry, therefore it has potential problems with both $K$-$\bar{K}$ and $D$-$\bar{D}$ mixings which can not be protected by simple $U(1)_d$ horizontal asymmetries~\cite{Csaki:2008eh}.}
\begin{align}
c_{u_L}&=c_{d_L} \simeq 0.44, &c_{u_R}&=c_{d_R} \simeq 0.80, \nonumber\\
c_{c_L}&=c_{s_L} \simeq 0.62, &c_{c_R} &\simeq 0.62, &c_{s_R} &\simeq 0.49 \,,
\nonumber \\
c_{t_L}&=c_{b_L} \simeq 0.51, &c_{t_R}&\simeq -1.30, &c_{b_R}&\simeq 0.53\,.
\label{cset1}
\end{align}
The $\AFBtt$ as a function of $m_{t \bar{t}}$ for both scenarios is presented in Figs. \ref{fig:afb} and \ref{partonic} which show a similar bending behavior as the original axigluon model (see Fig.~\ref{Fig:afbmtt2}). Similar results have been found in an RS Higgsless model \cite{Barcelo:2011fw}.

%% file: afb_mtt2.tex
\setlength{\unitlength}{0.240900pt}
\ifx\plotpoint\undefined\newsavebox{\plotpoint}\fi
\sbox{\plotpoint}{\rule[-0.200pt]{0.400pt}{0.400pt}}%
\begin{picture}(1500,900)(270,0)
\font\gnuplot=cmr10 at 12pt
\gnuplot
\sbox{\plotpoint}{\rule[-0.200pt]{0.400pt}{0.400pt}}%
\put(504.0,160.0){\rule[-0.200pt]{4.818pt}{0.400pt}}
\put(479,160){\makebox(0,0)[r]{-0.2}}
\put(1175.0,160.0){\rule[-0.200pt]{4.818pt}{0.400pt}}
\put(504.0,259.0){\rule[-0.200pt]{4.818pt}{0.400pt}}
\put(479,259){\makebox(0,0)[r]{-0.1}}
\put(1175.0,259.0){\rule[-0.200pt]{4.818pt}{0.400pt}}
\put(504.0,357.0){\rule[-0.200pt]{4.818pt}{0.400pt}}
\put(479,357){\makebox(0,0)[r]{ 0}}
\put(1175.0,357.0){\rule[-0.200pt]{4.818pt}{0.400pt}}
\put(504.0,456.0){\rule[-0.200pt]{4.818pt}{0.400pt}}
\put(479,456){\makebox(0,0)[r]{ 0.1}}
\put(1175.0,456.0){\rule[-0.200pt]{4.818pt}{0.400pt}}
\put(504.0,554.0){\rule[-0.200pt]{4.818pt}{0.400pt}}
\put(479,554){\makebox(0,0)[r]{ 0.2}}
\put(1175.0,554.0){\rule[-0.200pt]{4.818pt}{0.400pt}}
\put(504.0,653.0){\rule[-0.200pt]{4.818pt}{0.400pt}}
\put(479,653){\makebox(0,0)[r]{ 0.3}}
\put(1175.0,653.0){\rule[-0.200pt]{4.818pt}{0.400pt}}
\put(504.0,751.0){\rule[-0.200pt]{4.818pt}{0.400pt}}
\put(479,751){\makebox(0,0)[r]{ 0.4}}
\put(1175.0,751.0){\rule[-0.200pt]{4.818pt}{0.400pt}}
\put(504.0,850.0){\rule[-0.200pt]{4.818pt}{0.400pt}}
\put(479,850){\makebox(0,0)[r]{ 0.5}}
\put(1175.0,850.0){\rule[-0.200pt]{4.818pt}{0.400pt}}
\put(553.0,160.0){\rule[-0.200pt]{0.400pt}{4.818pt}}
\put(553,110){\makebox(0,0){ 400}}
\put(553.0,830.0){\rule[-0.200pt]{0.400pt}{4.818pt}}
\put(652.0,160.0){\rule[-0.200pt]{0.400pt}{4.818pt}}
\put(652,110){\makebox(0,0){ 500}}
\put(652.0,830.0){\rule[-0.200pt]{0.400pt}{4.818pt}}
\put(751.0,160.0){\rule[-0.200pt]{0.400pt}{4.818pt}}
\put(751,110){\makebox(0,0){ 600}}
\put(751.0,830.0){\rule[-0.200pt]{0.400pt}{4.818pt}}
\put(850.0,160.0){\rule[-0.200pt]{0.400pt}{4.818pt}}
\put(850,110){\makebox(0,0){ 700}}
\put(850.0,830.0){\rule[-0.200pt]{0.400pt}{4.818pt}}
\put(948.0,160.0){\rule[-0.200pt]{0.400pt}{4.818pt}}
\put(948,110){\makebox(0,0){ 800}}
\put(948.0,830.0){\rule[-0.200pt]{0.400pt}{4.818pt}}
\put(1047.0,160.0){\rule[-0.200pt]{0.400pt}{4.818pt}}
\put(1047,110){\makebox(0,0){ 900}}
\put(1047.0,830.0){\rule[-0.200pt]{0.400pt}{4.818pt}}
\put(1146.0,160.0){\rule[-0.200pt]{0.400pt}{4.818pt}}
\put(1146,110){\makebox(0,0){ 1000}}
\put(1146.0,830.0){\rule[-0.200pt]{0.400pt}{4.818pt}}
\put(504.0,160.0){\rule[-0.200pt]{0.400pt}{166.221pt}}
\put(504.0,160.0){\rule[-0.200pt]{166.462pt}{0.400pt}}
\put(1195.0,160.0){\rule[-0.200pt]{0.400pt}{166.221pt}}
\put(504.0,850.0){\rule[-0.200pt]{166.462pt}{0.400pt}}
\put(329,505){\makebox(0,0){$A^{t}_{FB}$}}
\put(849,35){\makebox(0,0){$M_{t\bar{t}} \text{(GeV)}$ Edge}}
\put(1000,805){\makebox(0,0)[r]{Below $M_{t\bar{t}}$ Edge}}
\put(553,370){\raisebox{-.8pt}{\makebox(0,0){$\Diamond$}}}
\put(603,384){\raisebox{-.8pt}{\makebox(0,0){$\Diamond$}}}
\put(652,397){\raisebox{-.8pt}{\makebox(0,0){$\Diamond$}}}
\put(701,408){\raisebox{-.8pt}{\makebox(0,0){$\Diamond$}}}
\put(751,418){\raisebox{-.8pt}{\makebox(0,0){$\Diamond$}}}
\put(800,426){\raisebox{-.8pt}{\makebox(0,0){$\Diamond$}}}
\put(850,432){\raisebox{-.8pt}{\makebox(0,0){$\Diamond$}}}
\put(899,436){\raisebox{-.8pt}{\makebox(0,0){$\Diamond$}}}
\put(948,439){\raisebox{-.8pt}{\makebox(0,0){$\Diamond$}}}
\put(998,441){\raisebox{-.8pt}{\makebox(0,0){$\Diamond$}}}
\put(1047,443){\raisebox{-.8pt}{\makebox(0,0){$\Diamond$}}}
\put(1096,444){\raisebox{-.8pt}{\makebox(0,0){$\Diamond$}}}
\put(1146,445){\raisebox{-.8pt}{\makebox(0,0){$\Diamond$}}}
\put(1085,805){\raisebox{-.8pt}{\makebox(0,0){$\Diamond$}}}
\put(1000,755){\makebox(0,0)[r]{Above $M_{t\bar{t}}$ Edge}}
\put(553,472){\makebox(0,0){$+$}}
\put(603,517){\makebox(0,0){$+$}}
\put(652,565){\makebox(0,0){$+$}}
\put(701,608){\makebox(0,0){$+$}}
\put(751,637){\makebox(0,0){$+$}}
\put(800,646){\makebox(0,0){$+$}}
\put(850,627){\makebox(0,0){$+$}}
\put(899,582){\makebox(0,0){$+$}}
\put(948,514){\makebox(0,0){$+$}}
\put(998,436){\makebox(0,0){$+$}}
\put(1047,357){\makebox(0,0){$+$}}
\put(1096,285){\makebox(0,0){$+$}}
\put(1146,223){\makebox(0,0){$+$}}
\put(1085,755){\makebox(0,0){$+$}}
\put(504.0,160.0){\rule[-0.200pt]{0.400pt}{166.221pt}}
\put(504.0,160.0){\rule[-0.200pt]{166.462pt}{0.400pt}}
\put(1195.0,160.0){\rule[-0.200pt]{0.400pt}{166.221pt}}
\put(504.0,850.0){\rule[-0.200pt]{166.462pt}{0.400pt}}
\end{picture}

%% file: 5lhc.tex
\section{LHC signatures}
\label{sec:LHC}

Since the LHC is a proton-proton collider there is no universally forward direction that one can define. However, because at large $x$ the 
the PDFs are dominated by the valence quarks, there are more top quarks produced in the forward region, while there is a slight excess of antitops in the central region already in the SM. This observation motivates several definitions of charge asymmetries at LHC (see \cite{Kuhn:1998jr,Kuhn:1998kw,Antunano:2007da,Wang:2010tg,Xiao:2011kp,AguilarSaavedra:2011vw,Hewett:2011wz,Krohn:2011tw,Bai:2011uk}).  For instance, one can define the differential lepton charge asymmetry
\bea
A^C_{pp}(y) = \frac{\frac{\displaystyle dN(Q)}{\displaystyle dy}
- \frac{\displaystyle dN(\bar{Q})}{\displaystyle dy}}
{\frac{\displaystyle dN(Q)}{\displaystyle dy} 
+ \frac{\displaystyle dN(\bar{Q})}{\displaystyle dy}}~,
\eea
from the semileptonic top decays as a function of rapidity 

\begin{figure}[t]
	\centering
	\vspace*{.5cm}
			\includegraphics[width=0.43\textwidth,angle=0]{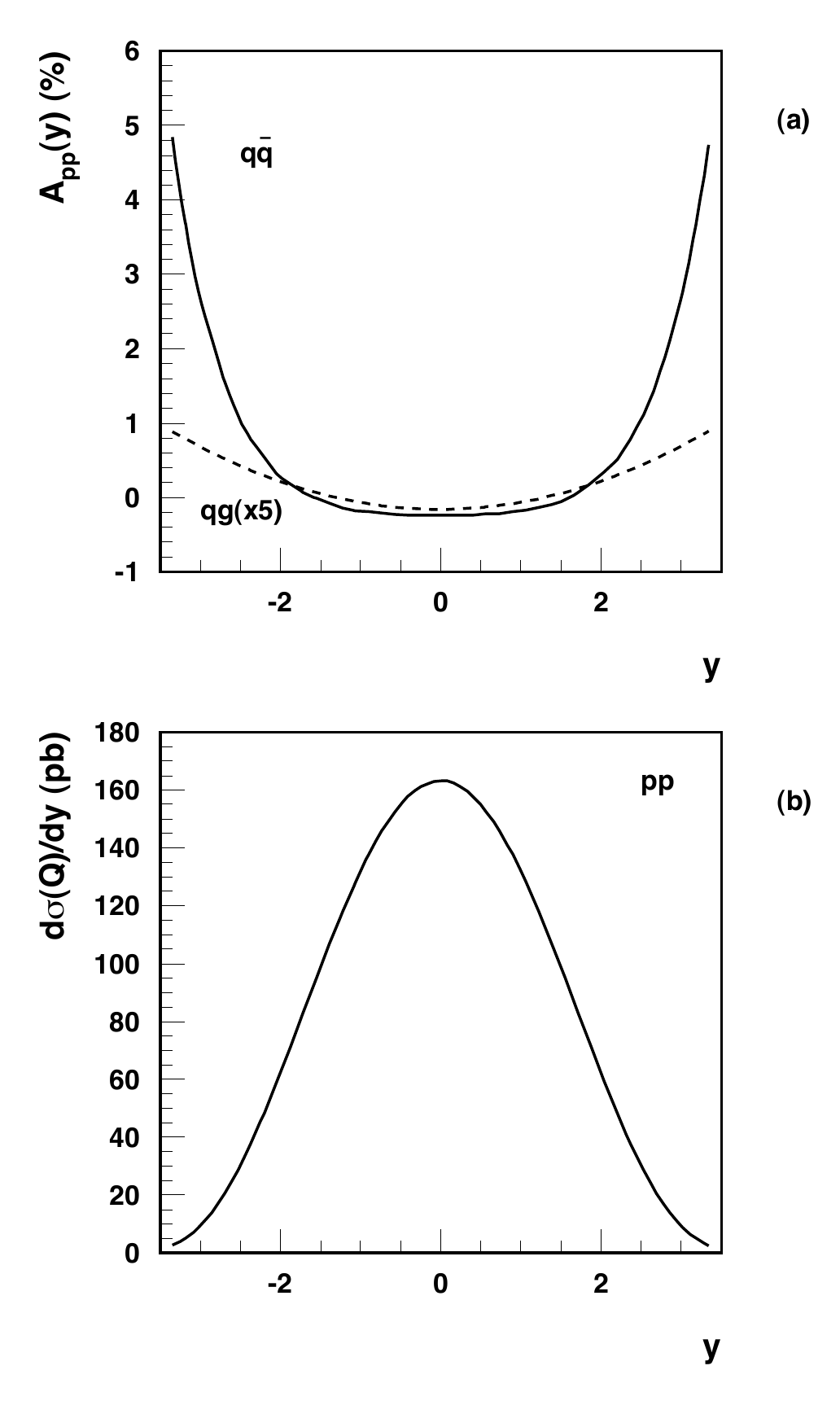}
\caption{\label{fig:AFBrapidity} \small{Rapidity distribution of charge asymmetry (a) and total cross section at Born order (b) of top quark production in proton-proton collisions, $\sqrt{s} =$ 14 TeV and $\mu = m_t$.} Contributions from $q \bar{q}$ fusion and flavor excitation, $gg(\bar{q}g)$, are shown separately. In the laboratory frame using CTEQ-1 PDFs. Taken from Ref.~\cite{Kuhn:1998kw}.  }
\end{figure}

Prediction for SM top quark
production at the LHC ($\sqrt{s}=14$~TeV) is shown in Fig.~\ref{fig:AFBrapidity}.
A sizable charge asymmetry is predicted in the large rapidity region. Therefore, there are two basic ways of defining integrated charge asymmetries. One focuses on the central region, which has a statistically much larger top sample while the charge asymmetry itself is small~\cite{Kuhn:1998jr,Kuhn:1998kw}.  The central charge asymmetry is thus defined as 
\beq
A_C^C(y_C)=\frac{N_t(|y|<y_C)-N_{\bar t}(|y|<y_C)}{N_t(|y|<y_C)+N_{\bar t}(|y|<y_C)},
\eeq
where $N_t$ ($N_{\bar t}$) is the number of events  where top (antitop)  has $|y|$ less then an appropriately chosen value $y_C$ that delineates the central from the forward region of the detector. Similarly one can define a forward (or "edge") charge asymmetry~\cite{Wang:2010tg,Xiao:2011kp}
\beq
A_C^F(y_C)=\frac{N_t(|y|>y_C)-N_{\bar t}(|y|>y_C)}{N_t(|y|>y_C)+N_{\bar t}(|y|>y_C)},
\eeq
which counts the difference between the number of events with top and antitop in the forward region. The event rate in this extreme region is much lower, but the asymmetry is large. It is interesting to note that in this region where the charged leptons are moving close to the beam line direction, the LHCb experiment may have a good potential to observe it~\cite{Kagan:2011yx}. Recently, there are also attempts to improve the measurements by using leptonic top decays in the central region and hadronic top decays in the forward region~\cite{Arguin:2011xm}.  

At present ATLAS and CMS measure a charge asymmetry defined as
\beq
A_C^\eta=\frac{N(|\eta_t|>|\eta_{\bar t}|)-N(|\eta_t|<|\eta_{\bar t}|)}{N(|\eta_t|>|\eta_{\bar t}|)+N(|\eta_t|<|\eta_{\bar t}|)},
\eeq
with $\eta_t$ and $\eta_{\bar t}$ being the pseudorapidities of the top and the antitop, respectively. Similarly, $A_C^y$ is defined by  replacing $\eta$ with $y$.
Very recently, the CMS updated its measurement of the charged asymmetry~\cite{newCMScharge} based on both rapidity and pseudo-rapidity distributions
\begin{align}\label{CMS-measurement}
A_C^\eta =& -0.016 \pm 0.030 (\textrm{stat.})^{+0.010}_{-0.019}(\textrm{syst.}),\\
A_C^y =& -0.013 \pm 0.026 (\textrm{stat.}) ^{+0.025}_{-0.020}(\textrm{syst.}),
\end{align}
to be compared with the SM predictions
\begin{align}
A_C^\eta(\textrm{SM}) = 0.013 \pm 0.001,\\
A_C^y(\textrm{SM}) = 0.011 \pm 0.001.
\end{align}
ATLAS also presented a measurement~\cite{AcEPSATLAS}
\beq\label{ATLAS-measurement}
A_C^y=-0.023 \pm 0.015 (\textrm{stat.}) \pm0.021(\textrm{syst.}),
\eeq
to be compared with MC@NLO prediction for SM
\beq
A_C^y=0.005\pm 0.001(\textrm{syst.}).
\eeq
Within the errors the latest measurements agree with the SM, while we discuss the expectations for NP models below. Before we proceed, let us point out the impressive improvement these measurements represent compared to the previous CMS result~\cite{CMScharge}
\bea
A_C^\eta = 0.060 \pm 0.134 (\textrm{stat.}) \pm 0.026 (\textrm{syst.}) \ .
\eea

\begin{figure}[bt]
	\centering
			\includegraphics[width=0.44\textwidth,angle=0]{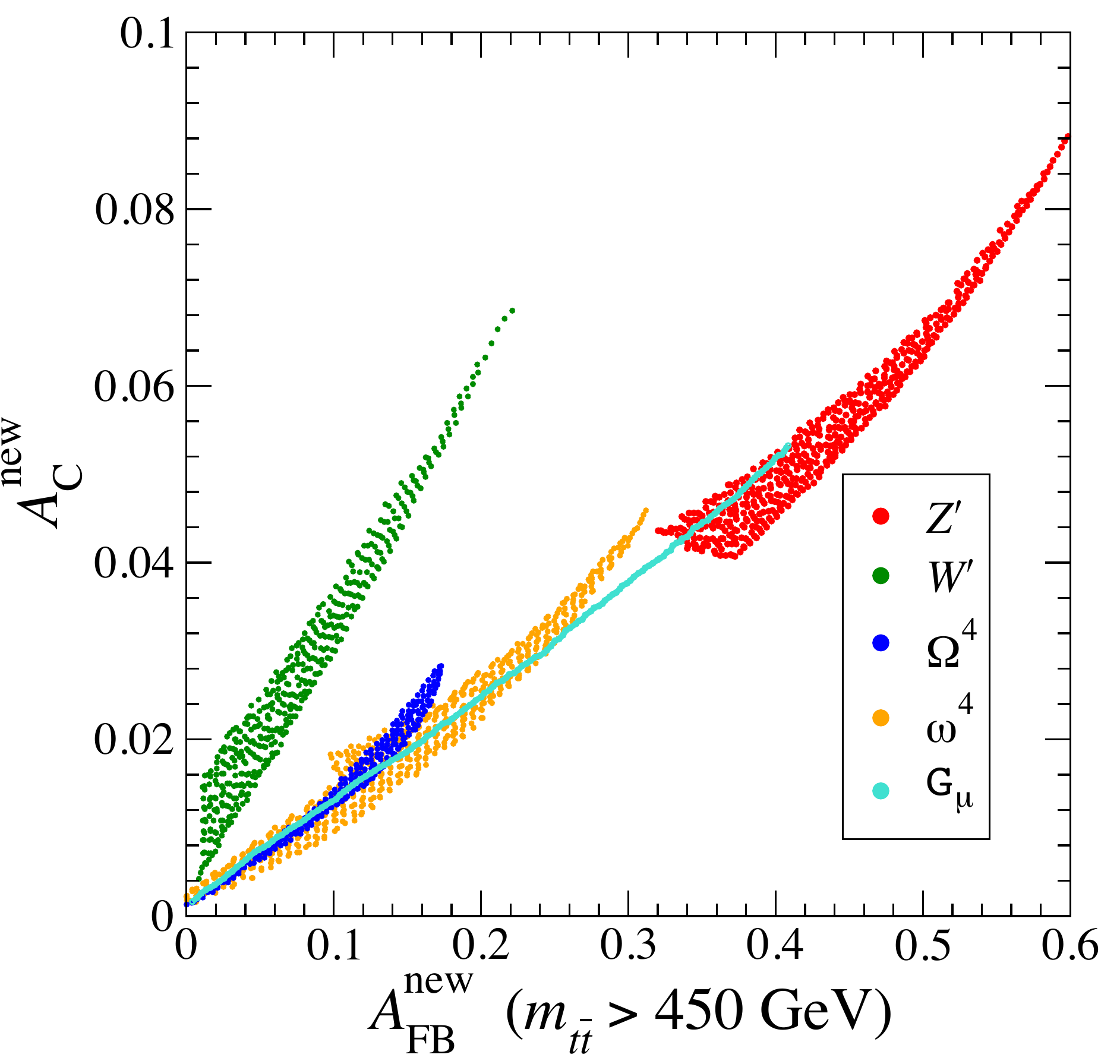}
\caption{\label{fig:Ac} \small{Allowed regions for NP contributions to the $\AFBtt$ at the Tevatron ($\AFB^{\rm new} \equiv \Ahigh - \AFB^{\rm h, SM}$) and the inclusive charge asymmetry at the LHC ($A_C^{\rm new} \equiv A_C^\eta - A_C^{\eta,\rm SM} $).} The $Z'$, $W'$, $\Omega_4$, $\omega_4$ and $G_\mu$ represent the $t$-channel $Z'$, $W'$, color sextet scalar, color triplet scalar and the $s$-channel axigluon model, respectively. Taken from Ref. \cite{AguilarSaavedra:2011hz}. }
\end{figure}

Figure. \ref{fig:Ac} shows results of a comprehensive parameter scan for a number of popular NP models used to explain $\AFBtt$ measurements at CDF. There is a clear correlation between a large asymmetry in $\AFBtt$ for $m_{t \bar{t}} > 450$~GeV ($\Ahigh$) and a sizable $A_C^\eta$ of a few percent predicted at the LHC. The slope for the $W'$ model (green dots) is larger because of larger enhancement of the $d \bar{d}$ initial state PDF at the LHC. Some of the points for models on Figure. \ref{fig:Ac} are already in tension with the ATLAS and CMS measurements for  $A_C^\eta$~\eqref{CMS-measurement},~\eqref{ATLAS-measurement}. With further improvements we should be able to reach a concrete answer about the top charge asymmetry measurements at the Tevatron and LHC.    

One byproduct of models which explain $\AFBtt$ is the existence of top resonances. There are two types of top resonance searches that can be motivated by these models. One is a search for $t \bar{t}$ resonance that arises in the $s$-channel models. In the $t$-channel models on the other hand one predicts a $t$+jet resonance in the $t\bar t$+jet channel~\cite{Shu:2009xf,Gresham:2011dg}. In this case the new particles (that for $t\bar t$ final state are exchanged in the $t$-channel) can be singly produced from the gluon-valence quark initial state. 

If tops originate from a decay of a heavy color octet particle $G'$ they are highly boosted. Their identification (or the top tagging) is more challenging experimentally than identifying tops in threshold production. If such a resonance is observed, further experimental information about its properties can be deduced. 
For instance, tops produced by the $s$-channel resonance are highly polarized due to a large axial-coupling. Therefore, one can use observables related to the charged lepton angular distributions to help with the identification. For detailed studies, see Refs.~\cite{Lillie:2007yh,Agashe:2006hk} on search strategies for the bulk RS KK gluons.

Searching for the top jet resonance at the LHC was proposed in the analyses of a $t$-channel color scalar models where the mass of the scalar is considerable larger than the mass of top~\cite{Shu:2009xf,Dorsner:2009mq}. In a more recent analysis, the polar decay angle of the $t$-channel particle was used to help the reconstruction when the invariant top+jet resonance has a large SM background. In the lab frame, the polar angle between the reconstructed (anti)top and the remaining jet is collimated in the same direction as the light $t$-channel particle and it can be used as an efficient cut in the low mass region~\cite{Gresham:2011dg}.
Finally, it has recently been pointed out that a fraction of on-shell production of light (top+jet) resonances might pass experimental cuts of the $t\bar t$ production measurements at the LHC, thus effectively enhancing the predicted $t\bar t$ cross-section in $t$-channel models with light NP particles~\cite{Gresham-t}. Thus such models might already be in some tension with the existing LHC $t\bar t$ cross-section and $m_{t\bar t}$ spectrum measurements~\cite{ATLAS-CONF-2011-087,ATLAS-CONF-2011-108,CMS-PAS-TOP-11-001}.

The $\AFBtt$ has an intrinsic connection with the top polarization and top spin correlations. The reason is that the top polarization is the one of the two possible sources for a nonzero $\AFBtt$ at the Tevatron (the other one being the Rutherford enhancement peak in the $t$-channel process). In all the $s$-channel and $t$-channel di-quark models, it is the top polarization that provides the main source of the $\AFBtt$ \footnote{In the $t$-channel models, the $\AFBtt$ provided from top polarization will enhance (compete with) the one from Rutherford enhancement in the vector boson (scalar) models~\cite{Shu:2009xf}.} 

At the LHC, the top polarization or the top spin correlations can be used either to distinguish among different models or to reduce the large SM $t \bar{t}$ background~\cite{Hewett:2011wz,Bai:2011uk,Krohn:2011tw}, in particular the large additional contributions from the $gg$ initial state. For the NP model discrimination there are two useful observables defined in the helicity basis
\begin{align}
\label{eq:thetaasymmlhc}
{\cal P}_ {n} =&\frac{N(\cos\theta_{\ell, n} > 0) - N(\cos\theta_{\ell, n} < 0))  } {N(\cos\theta_{\ell, n} > 0) + N(\cos\theta_{\ell, n} < 0)) },\\
{\cal A}^\ell_ {c_1c_2} =&\frac{N(c_1c_2> 0) - N(c_1c_2< 0))  } {N(c_1c_2 > 0) + N(c_1c_2< 0))  },
\end{align}
where $n$ is given by the direction of the parent topÕs momentum in the CM frame in the helicity basis and $c_1=\cos\theta_{\ell_1, n} $, $c_2=\cos\theta_{\ell_2, n} $.
Expectations for ${\cal P}_ {n}$ and ${\cal A}^\ell_ {c_1c_2}$ for a set of benchmark models (with details given in the Appendix B of Ref. \cite{Krohn:2011tw}) are tabulated in Table \ref{tab:helicitylhc} and Table \ref{tab:helicitycorr}, respectively.
Judging from the benchmark models  top polarization ${\cal P}_ {n}$ appears to be more sensitive to NP effects than the top spin correlation ${\cal A}^\ell_ {c_1c_2}$. It is interesting to note that the $W'$ models look very promising with a large predicted signal due to large enhancement of the $d \bar{d}$ initial state PDF going from Tevatron to the LHC. 

\begin{table*}
\begin{center}
\begin{tabular}{c|cccccc}
\hline\hline
& $G_A ($\%$) $ & $G_L ($\%$) $ & $G_R ($\%$) $ & $W'(\%) $& SM($\%$)   \\ 
Selection cuts  & 1 &	-1 &	 4 &	18 &	1 $(\pm1.2)$ \\ 
$m_{t\bar t} > 450 ~{\rm GeV}$  & 2 &	-2 &	6 &	26 &	0  $(\pm1.7)$\\ 
 $|y(t)+y(\bar t)| > 2$ & 0 &	-4 &	3 &	19 &	-2  $(\pm3.2)$\\ 
 \hline\hline
\end{tabular}
\end{center}
\caption{Net polarization ${\cal P}_ {\rm h} $ in the helicity basis at the 7 TeV LHC for a set of NP models, 2 TeV axialgluons with axial (left-handed, right-handed) couplings to top $G_A (G_{L,R})$, and a $400$ GeV $W'$. In parenthesis in the last column are given the $1\sigma$ statistical uncertainties  assuming $5~{\rm fb}^{-1}$ of integrated luminosity.\label{tab:helicitylhc} Table from Ref. \cite{Krohn:2011tw}.}
\end{table*}

\begin{table*}
\begin{center}
\begin{tabular}{c|cccccc}
\hline\hline
& $G_A ($\%$) $ & $G_L ($\%$) $ & $G_R ($\%$) $ & $W'(\%) $& SM($\%$)   \\ \hline
Selection cuts  & -2 &	-3 &	-2 &	7 &	-4  $(\pm1.2)$\\ 
$m_{t\bar t} > 450 ~{\rm GeV}$  & 1 &	0 &	1 &	12 &	-2  $(\pm1.7)$\\ 
 $|y(t)+y(\bar t)| > 2$ & 3 &	 0 &	0 &	12 &	3  $(\pm3.2)$\\
 \hline\hline
\end{tabular}
\end{center}
\caption{Same as for Table~\ref{tab:helicitylhc}, but for spin correlation ${\cal A}^\ell_ {c_1c_2}$. 
\label{tab:helicitycorr}
Table from Ref.~\cite{Krohn:2011tw}.}
\end{table*}

%% file: 6conclusions.tex
\section{Conclusions}
\label{sec:Conclusions}
The models proposed to explain the large $\AFBtt$ measured  by CDF and D\O\ have observable consequences at the LHC. This is inevitable, since in order to influence $t\bar t$ production at Tevatron, the new fields need to couple both to light quarks and tops. A number of features can thus be expected, from a deviation in the charge asymmetries in $t\bar t$ production, to $t+$jet resonances or $t\bar t$ resonances. The final analyses at  Tevatron  are expected to use twice as much data as the present ones. Further experimental improvements both at the Tevatron and at the LHC will thus tell whether the effect is due to new physics and, if so, due to what kind of new physics.